\pdfoutput=1

\documentclass[
  twocolumn,
  prb,
  showpacs,
  amsmath,
  amssymb,
  superscriptaddress
]{revtex4}

\usepackage{bm}
\usepackage{graphicx}
\usepackage{amssymb}
\usepackage{color}
\usepackage{braket}
\usepackage{pifont}
\usepackage{empheq}
\usepackage{scalerel,stackengine}
\stackMath
\newcommand\reallywidehat[1]{%
\savestack{\tmpbox}{\stretchto{%
  \scaleto{%
    \scalerel*[\widthof{\ensuremath{#1}}]{\kern-.6pt\bigwedge\kern-.6pt}%
    {\rule[-\textheight/2]{1ex}{\textheight}}
  }{\textheight}%
}{0.5ex}}%
\stackon[1pt]{#1}{\tmpbox}%
}

\renewcommand{\v}[1]{\textbf{\textit #1}}
\newcommand{\sgn}[1]{\text{sgn}({#1})}

   \let\k=\kappa

\def\bpm{\begin{pmatrix}}
\def\epm{\end{pmatrix}}
\def\be{\begin{equation}}
\def\ee{\end{equation}}
\def\bea{\begin{eqnarray}}
\def\eea{\end{eqnarray}}
\def\ba{\begin{array}}
\def\ea{\end{array}}

\def\vep{{\varepsilon}}

\def\bfr{\mathbf{r}}

\newcommand{\bk}{{\bf k}}

\newcommand{\bp}{{\bf p}}

\newcommand{\mathsym}[1]{{}}
\newcommand{\unicode}[1]{{}}

\newcommand{\bd}{{\bf d}}
\newcommand{\bsig}{{\boldsymbol{\sigma}}}
\newcommand{\bK}{{\bf K}}
\newcommand{\bfa}{{\bf a}}

\begin{document}

\title{Photogalvanic effect in Weyl semimetals}

\author{E. J.\ K\"onig}
\affiliation{Department of Physics and Astronomy, Rutgers University, Piscataway, New Jersey, 08854, USA}
\affiliation{Department of Physics, University of Wisconsin-Madison, Madison, Wisconsin 53706, USA}

\author{H.-Y. Xie}
\affiliation{Department of Physics, University of Wisconsin-Madison, Madison, Wisconsin 53706, USA}

\author{D. A. Pesin}
\affiliation{Department of Physics and Astronomy, University of Utah, Salt Lake City, UT 84112 USA}

\author {A.  Levchenko} 
\affiliation{Department of Physics, University of Wisconsin-Madison, Madison, Wisconsin 53706, USA}

\date{\today}

\begin{abstract}
We theoretically study the impact of impurities on the photogalvanic effect (PGE) in Weyl semimetals with weakly tilted Weyl cones. Our calculations are based on a two-nodes model with an inversion symmetry breaking offset and we employ a kinetic equation approach in which both optical transitions as well as particle-hole excitations near the Fermi energy can be taken into account. We focus on the parameter regime with a single photoactive node and control the calculation in small impurity concentration. Internode scattering is treated generically and therefore our results allow to continuously interpolate between the cases of short range and long range impurities. We find that the time evolution of the circular PGE may be nonmonotonic for intermediate internode scattering. Furthermore, we show that the tilt vector introduces three additional linearly independent components to the steady state photocurrent. Amongst them, the photocurrent in direction of the tilt takes a particular role inasmuch it requires elastic internode scattering or inelastic intranode scattering to be relaxed. It may therefore be dominant. The tilt also generates skew scattering which leads to a current component perpendicular to both the incident light and the tilt. We extensively discuss our findings and comment on the possible experimental implications.
\end{abstract}


\maketitle

\section{Introduction}

In the present days, Weyl semimetals [\onlinecite{Burkov2016}] enjoy significant scientific interest which is generated by a multitude of reasons. From the theoretical viewpoint, this class of materials is special since it represents a solid state realization of Weyl's theory of chiral relativistic fermions [\onlinecite{Weyl1929}] -- an elegant theory which regrettably seems to have lost its connection to fundamental particle physics after the discovery of neutrino oscillations [\onlinecite{NeutrinoOscillations}]. In the context of transport theories, the unique features related to the chiral anomaly [\onlinecite{Bertlmann2000}] are at the basis [\onlinecite{SonSpivak2013}] of the experimentally observed giant magnetoresistance [\onlinecite{XiongOng2015}]. From the viewpoint of applications, Weyl semimetals are attractive as their topologically protected band touching associated with a Berry curvature singularity may allow to explore and exploit novel regimes and phenomena of semiconductor physics. As we understand now, the appearance of Weyl nodes in systems lacking inversion or time reversal symmetry is far from being exceptional and the observation of Weyl physics has been already reported in various materials [\onlinecite{WeylRealizations}]. One particularly exciting phenomenon occurring in Weyl semimetals is the photogalvanic effect (PGE) [\onlinecite{ChanLee2016, WuOrenstein2016, MorimotoMoore2016, Jarillo-Herrero-Gedik2017}], i.e. the generation of current due to the exposure to light. In contrast to most ordinary semiconductors, Weyl semimetals are susceptible to lowest frequencies and allow for novel technologies in the infrared. This consequence of the protected gapless spectrum comes along with the theoretical prediction of a quantized circular PGE [\onlinecite{deJuanMoore2017}] which is awaiting its experimental verification. 

\begin{figure}
\includegraphics[scale=.45]{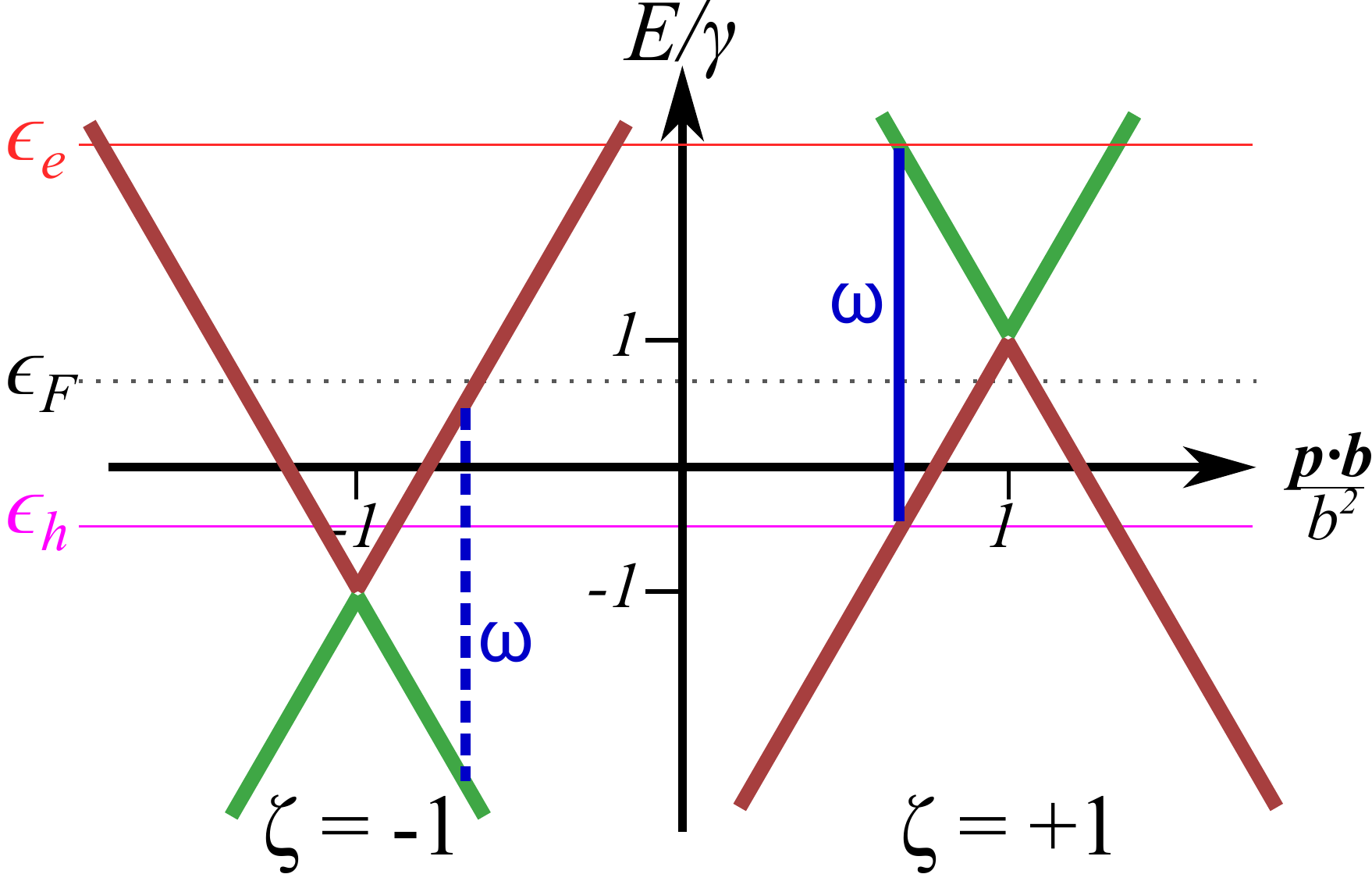} 
\caption{[Color online] Cut through the spectrum of $H_{\rm kin}$ defined by Eq.~\eqref{eq:Hkin} for the case $\hat t \perp \v b$. The green [red] dispersion relation represents bands with negative [positive] Berry flux $\text{sgn}(\widehat{\v p - \zeta \v b} \cdot \boldsymbol{\Omega}_\xi) = -\xi = -1$ [$\text{sgn}(\widehat{\v p - \zeta \v b} \cdot \boldsymbol{\Omega}_\xi) = -\xi = 1$]. When $0 < \epsilon_F < \gamma$ and $\frac{\omega}{2} \in (\gamma - \epsilon_F, \gamma + \epsilon_F)$, optical transitions occur in the $\zeta = +1$ cone, only (solid vertical line), while transitions in the $\zeta = -1$ cone are Pauli-blocked (dashed vertical line). In this plot $\omega = 3 \gamma$.}
\label{fig:Opttransition}
\end{figure}

The theoretical foundations of PGE go back to the seventies (for review see Refs.~[\onlinecite{BelinicherSturman1980,SturmanFridkin1992,Ivchenko2005}]) while modern derivations rely on the nonlinear susceptibility framework [\onlinecite{SipeShkrebtii2000}], Floquet theory [\onlinecite{MorimotoNagaosa2016}] and the Keldysh quantum kinetic equation approach [\onlinecite{Koenig2017}]. In this paper, we concentrate on the dc response. It is common to distinguish two contributions: the \textit{injection} current and the \textit{shift} current. The former of the two represents the current generated by photoelectrons which are excited by optical (vertical) transitions with an anisotropic transition rate. In the steady state, this contribution diverges linearly when the current relaxation time becomes infinite. Differently stated, in the absence of a relaxation mechanism, the injection current grows linearly in time. In contrast, the shift current [\onlinecite{vonBaltzKraut1981,BelinicherSturman1982}] is due to the real space displacement that electrons undergo upon an optical transition. This contribution is always finite and thus subleading for long relaxation times.  

Therefore, in the present paper we keep only the dc injection current. We focus on the parameter regime with a single photoactive node which materializes the predicted topological quantization [\onlinecite{deJuanMoore2017}]. We investigate its relaxation due to impurities taking into account both intra and inter valley scattering in a model of two Weyl nodes. We scrutinize the time evolution of the current in response to a sudden illumination as well as its steady state characteristics. Furthermore we study the effect of a weak tilt in the Weyl spectrum. As we show, the interplay of impurity scattering and tilt leads to skew scattering and to the appearance of current components transversal to the direction of illumination. 

It is also worthwhile to emphasize, that the current injection mechanism for the nonlinear PGE response at large frequencies is strongly different from the nonlinear intraband response at small frequencies considered in Refs.~[\onlinecite{SodemannFu2015,ZyuzinZyuzin2017,IshizukaNagaosa2017}]. We explicitly show that the latter exactly vanishes in our model for the case of absent tilt and intervalley scattering and that it is generically subdominant for frequencies exceeding the elastic scattering rate.

This article is structured as follows. In Sec.~\ref{sec:Model} we introduce the model under consideration and the assumptions for the calculation. In Sec.~\ref{sec:Calculation} we present a sketch of the technical derivation as well as the main results, which are subsequently discussed in Sec.~\ref{sec:Discussion}. We conclude with a summary and an outlook about the experimental relevance of our findings, Sec.~\ref{sec:outlook}. The Appendix contains details on the model (Appendix \ref{app:model}), the elastic scattering rates (Appendix \ref{app:Rates}), the relaxation of the photocarriers (Appendix \ref{app:Relaxation}), the intraband response (Appendix \ref{app:intraband}), and a microscopic tight binding model (Appendix \ref{app:MicroModel}).

\section{Model and Assumption}
\label{sec:Model}

The terms entering the free Hamiltonian $H_0(\v p, \v x) = H_{\rm kin}(\v p) + H_{\rm dis}(\v x)$ are as follows. The kinetic term is a generalized low energy version of lattice models presented in Refs.~[\onlinecite{ShapourianHughes2016,deJuanMoore2017}] and Appendix~\ref{app:MicroModel}
\begin{subequations}
\begin{equation}
H_{\rm kin}(\v p) = 
\sum_{\zeta = \pm} \zeta [ v (\v p - \zeta \v b)  \cdot \boldsymbol \sigma + u \hat t \cdot (\v p - \zeta \v b)+  \gamma ] \frac{1 +\zeta \kappa_3}{2}. \label{eq:Hkin}
\end{equation}
Here, Pauli matrices $\boldsymbol{\sigma} = (\sigma_1, \sigma_2, \sigma_3)$ act in spin space, while Pauli matrices $\boldsymbol{\kappa} = (\kappa_1, \kappa_2, \kappa_3)$ act in node space. The terms proportional to velocities $v, u$ with $0 \leq u \ll v$ represent Weyl dispersion and the tilt in direction $\hat t$ respectively, the energy scale $\gamma$ is the offset between the two nodes. The spectrum $\epsilon_{\xi\zeta}(\v p) = \zeta (\xi v \vert \v p - \zeta \v b \vert + u \hat t\cdot [\v p - \zeta \v b] + \gamma) $ of the kinetic Hamiltonian is characterized by quantum numbers $\zeta = \pm 1$ (node), $\xi = \pm 1$ (band) and $\v p$ (momentum) and plotted in Fig.~\ref{fig:Opttransition}. 
This kinetic term is supplemented by a disorder potential 
\begin{equation}
H_{\rm dis}(\v x) = V(\v x) [\mathbf 1_\kappa + \kappa_1], \label{eq:Hdis}
\end{equation}
where $V(\v x) = \sum_{\v R_j} \mathcal{V}(\v x- \v R_j)$ is a sum over impurity potentials $\mathcal{V}(\v x)$ which are centered at uniformly distributed positions $\v R_j$ in $\mathbb R^3$.
Finally there is a homogeneous electric field $e \v E(t) = e \sum_{\pm} \boldsymbol{\mathcal E}_{\pm}e^{\pm i \omega t} = - \nabla \Phi(\v x,t)$ enclosed in our model Hamiltonian by means of $H = H_{0}(\v p, \v x) + \Phi(\v x, t)$ ($e$ is the elementary charge).
\label{eq:H}
\end{subequations}

By construction, the model defined by Eqs.~\eqref{eq:H} is not invariant under time reversal, space inversion or rotation symmetries, see App.~\ref{app:model} for more details. We note that the tilt direction could also be assumed to be equal in both nodes [\onlinecite{DetassisGrubinskas2017}]. In contrast, we chose a model with opposite tilt in opposite cones, so that the tilt preserves inversion symmetry.

Our calculation of the photocurrent in Weyl semimetals relies on the following simplifying assumptions.  First, we neglected any spatial dependence of the electric field which is justified for $v/c \rightarrow 0$, where $c$ is the speed of light. In doing so, we omit the photon drag effect. The assumption of a homogeneous electric field treatment within a 3D bulk Hamiltonian also relies on the second assumption of a long Thomas-Fermi screening length. In this limit, the contribution of surface states can be expected to be subdominant. Third, in neglecting all other sources of scattering we assume impurities to be the dominant source of photocurrent relaxation. While this statement is generally expected to be true at sufficiently low temperatures where electron-phonon and electron-electron scattering rates are small, we will explicitly show the limitations of this picture. Fourth, the current relaxation is calculated using a semiclassical kinetic equation approach which is justified if the mean free path of photoelectrons and photoholes exceeds their wavelength. This implies that the frequency of light is much larger than the elastic scattering rate. Formally, we implement this assumption by keeping only the leading $\mathcal O(n_{\rm imp}^{-1})$ terms in small impurity density. For the evaluation of quantum transition probabilities we truncate the T-matrix at next to leading (i.e. second) order in powers of weak impurity potentials. We also assume that the impurity potential is short ranged as compared to the wavelength of photocarriers (but it may be long-ranged as compared to $1/\vert \v b \vert$). Fifth, our calculation at the lowest temperatures is exponentially accurate if the energy difference between Fermi energy and the energy of photocarriers exceeds the temperature. Finally, all presented formulae are valid to linear order in $u/v \ll 1 $, unless stated otherwise.

\section{Sketch of the calculation and results}
\label{sec:Calculation}

In this section we present a sketch of the derivation of the photocurrent for a tilted, disordered Weyl semimetal. For the sake of a clearer presentation, we here restrict ourselves to the case of photocarrier generation at the $\zeta = +1$ cone only (see Fig.~\ref{fig:Opttransition}) and of absent intervalley scattering and relegate details and the more general case of finite intervalley scattering to Appendices~\ref{app:Rates} and \ref{app:Relaxation}. 

Our calculation is based on the Boltzmann kinetic equation [\onlinecite{LyandaGellerAndreev2015, DeyoSpivak2009}] describing the time evolution of the distribution function $f(\v p, t)$
\begin{equation}
\partial_t f(\v p, t) + \dot{\v p} \nabla_{\v p} f(\v p, t) = St_{\rm inj}[f] + St_{\rm dis}[f]. \label{eq:Boltzmann}
\end{equation}
Here, $\dot {\v p} = e \v E(t) $ and the collision integral contains two contributions. First, there is a term describing the excitation rates of photocarriers
\begin{subequations}
\begin{eqnarray}
St_{\rm inj} [ f ]&=&  2\pi \xi \zeta \delta (- \xi \zeta \omega + \epsilon_{\xi \zeta}(\v p) - \epsilon_{-\xi, \zeta}(\v p))  \notag \\
&& \Big \lbrace \frac{1}{4 p^2} \Big [ e\boldsymbol{\mathcal E}^T_+ (\mathbf 1 - \hat p \otimes \hat p) e\boldsymbol{\mathcal E}_{-}\Big ]\notag \\
&& + \frac{i}{2} \xi \zeta  (e\boldsymbol{\mathcal E}_{+} \times e \boldsymbol{\mathcal E}_{-}) \cdot  \boldsymbol{\Omega}_\xi \Big \rbrace. \label{eq:Sinj}
\end{eqnarray}
Here, we assumed states with energy $\epsilon_{\xi \zeta}(\v p)$ [$\epsilon_{-\xi \zeta}(\v p)$] to be empty [filled]. 
In the considered parameter range, $\xi = 1$ ($\xi = -1$) for photoelectrons (photoholes) and $\zeta = 1$, see Fig.~\ref{fig:Opttransition}. Clearly, terms proportional to the Berry curvature $\boldsymbol{\Omega}_\xi = -\xi {\hat p}/{2 p^2}$ change sign when the chirality of the photoactive node is reversed. Note that $St_{\rm inj}$ contains two anisotropic terms proportional to the direct product $\hat p \otimes \hat p$ and to $\boldsymbol{\Omega}_\xi$ as well as an isotropic term proportional to $\mathbf 1$. 

The second part of the collision integral describes the relaxation of photocarrieres by impurities
\begin{equation}
St_{\rm dis} [ f ] = - \int_{\v p} w_{\v p' \v p} f({\v p},t) - w_{\v p \v p'} f({\v p'},t).
\end{equation}
\end{subequations}
We introduced the shorthand notation $\int_{\v p} = \int d^3p/(2\pi)^3$.
Since impurity scattering is fully elastic, it may only relax anisotropic terms in the photoexcitation rates.
The intervalley scattering discussed below and in Appendix~\ref{app:Relaxation} may also relax the valley imbalance of the isotropic contribution to the photoexcitation rates.

In the limit $u/v = 0$, time reversal and rotational symmetries dictate that the scattering probability $w_{\v p' \v p}$ contains only terms proportional to $1$ and $\v p \cdot \v p'$ and thus $w_{\v p' \v p} = w_{\v p \v p'}$. This statement is fulfilled also when intervalley scattering is present, see Appendix \ref{app:Rates}. However, the symmetry of the scattering probability is lost in the presence of a finite tilt $u/v \neq 0$. Here we present the dominant contributions of symmetric and antisymmetric scattering probability:
\begin{subequations}
\begin{align}
w_{\v p', \v p}^{(s)} &\simeq  2 \pi \delta (\epsilon_{\xi \zeta} (\v p) - \epsilon_{\xi \zeta} (\v p') ) n_{\rm imp} \mathcal{V}^2_0 \frac{1 + \hat p \cdot \hat p'}{2}, \\
w_{\v p', \v p}^{(a)} & \simeq  2 \pi \delta (\epsilon_{\xi \zeta} (\v p) - \epsilon_{\xi \zeta} (\v p') ) \notag \\
&\times \left (-\frac{\pi u}{2v}\right ) \nu_\zeta(\epsilon_{\xi, \zeta}(\v p)) n_{\rm imp} \mathcal{V}_0^3 \hat t \cdot (\hat p \times \hat p'). \label{eq:SkewProb}
\end{align}
\end{subequations}
Here, $\mathcal{V}_{\v p}$ is the Fourier transform of the impurity potential and the density of states (DOS) is
\begin{equation}
\nu_{\zeta}(\epsilon) = \frac{(\zeta \epsilon - \gamma)^2}{2 \pi^2 v^3}.
\end{equation}
The skew scattering probability obtained in Eq.~\eqref{eq:SkewProb} relies on the third moment of the distribution function of the disorder potential. We note that, by means of both semiclassical [\onlinecite{SinitsynSinova2007}] and diffractive [\onlinecite{AdoTitov2015,KoenigLevchenko2016}] mechanisms, skew scattering may also occur for Gaussian disorder models. However, skew scattering probabilities due to Gaussian disorder contain an additional factor of $n_{\rm imp}$ and are thus subleading in the perturbation scheme employed here.

\begin{figure}
\includegraphics[scale=.65]{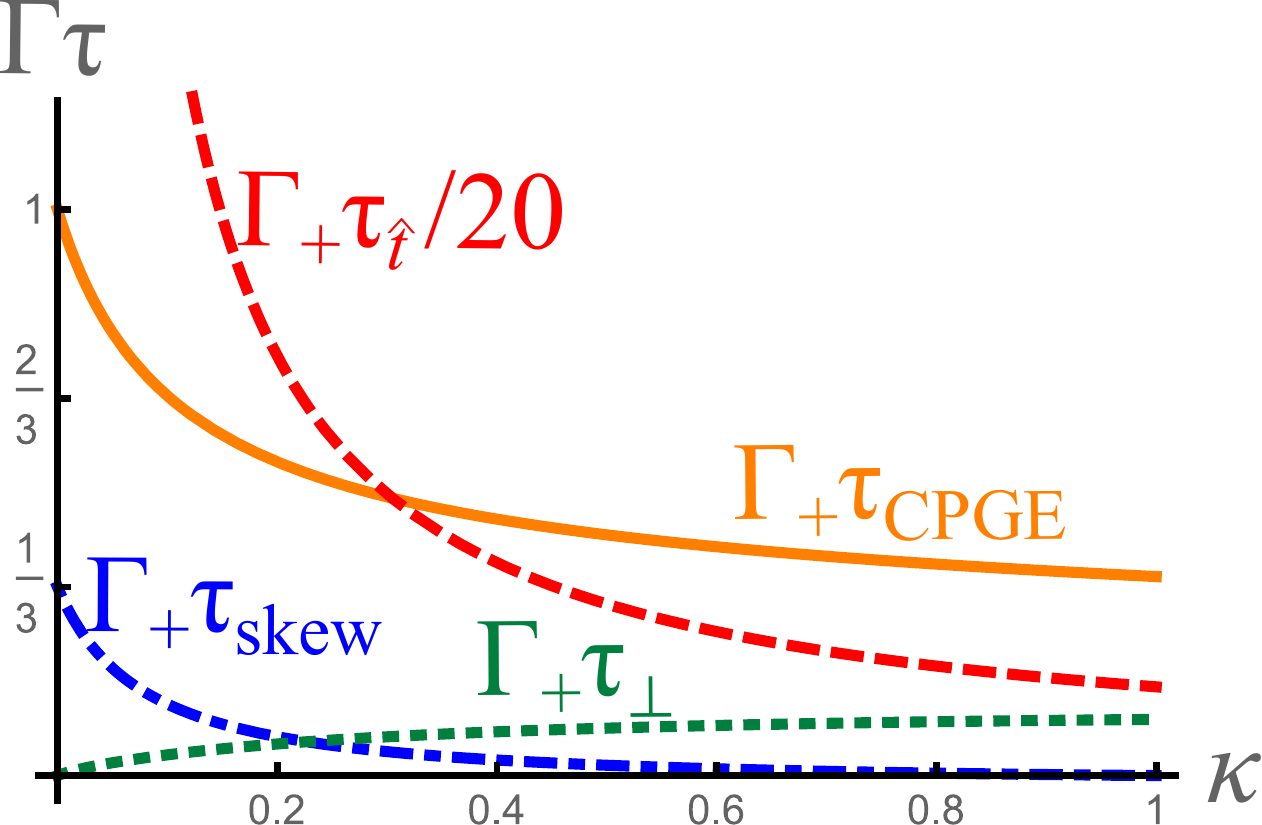} 
\caption{[Color online] Plot of the effective scattering times entering the final steady state solution of the photocurrent, Eq.~\eqref{eq:SteadyStatePGE}. As in Fig.~\ref{fig:Opttransition} we assumed $\omega = 3\gamma$ for this plot and furthermore we assumed $\Gamma_{\text{sk},+} = \Gamma_+/3$.}
\label{fig:PlotOfTaus}
\end{figure}

To obtain the photocurrent in the steady state, the solution for the distribution function can be obtained by equating the full collision integral to zero. We obtain the following nonequilibrium corrections to the Fermi-Dirac distribution function in band $\xi$
\begin{subequations}
\begin{eqnarray}
f_\xi(\v p) &=& \delta(\omega - 2 v p) \Gamma_+^{-1}\Bigg \lbrace \hat p_i I_i + \frac{2}{3} \hat p_i \hat p_j I_{ij} \notag \\
&&+\frac{u}{v} \xi \Big[\hat p_i \left (\hat t_i \frac{4 I}{5} - \frac{4 I_{ij} \hat t_j}{15}\right )  + \frac{2}{3}  \hat p_i \hat p_j \hat p_k \hat t_k I_{ij}\Big ]\notag \\
&& + \frac{u}{v}\epsilon_{ijk} \hat p_i I_j \hat t_k \frac{\Gamma_{sk,+}}{\Gamma_+} \Bigg \rbrace, \label{eq:deltaf}
\end{eqnarray}
where we introduced 
\begin{eqnarray}
&&I= \frac{2\pi v^2}{\omega^2} \xi e^2 \boldsymbol{\mathcal E}_+ \cdot \boldsymbol{\mathcal E}_- ,\\
&&I_i = \frac{2\pi v^2}{\omega^2} \xi \left \lbrace - i  e^2 ( \boldsymbol{\mathcal E}_+ \times \boldsymbol{\mathcal E}_-)_i\right\rbrace, \\
&&I_{ij} = \frac{2\pi v^2}{\omega^2} \xi  \left \lbrace -\text{Re}[e^2 \boldsymbol{\mathcal E}_{+,i} \boldsymbol{\mathcal E}_{-,j}]\right\rbrace .
\end{eqnarray}
\end{subequations}
We also introduced the intravalley scattering rates for momentum relaxation and skew scattering
\begin{subequations}
\begin{eqnarray}
&&\Gamma_\zeta (\epsilon)= \frac{2 \pi}{3} n_{\rm imp} \mathcal{V}_0^2 \nu_{\zeta} (\epsilon),\\
&&\Gamma_{\rm sk, \zeta} (\epsilon)= \frac{\pi^2}{3} n_{\rm imp} \mathcal{V}_0^3 \nu_{\zeta}^2 (\epsilon) ,
\end{eqnarray}
which both implicitly depend on the energy of the photocarriers by means of the density of states.
\end{subequations}
In Eq.~\eqref{eq:deltaf} the scalar part of the distribution function was omitted (it will be recovered shortly). Furthermore, Eq.~\eqref{eq:deltaf} was derived using the premise that only $\mathcal O(u/v)$ corrections proportional to odd powers of $\hat p$ enter the final expression for the photocurrent. 

We insert this expression into the definition of the current
\begin{equation}
\v j = - e \sum_{\xi = \pm 1} \int_{\v p}v \left [\xi \hat p + \frac{u}{v} \hat t \right ] f_{\xi}(\v p). \label{eq:CurrentDef}
\end{equation}
Here the sum over $\xi$ reflects the two types of photocarriers that the injection term generates: electrons in band $\xi = +1$ at energy $\epsilon_e = \gamma + \omega/2$ and holes in band $\xi = -1$ at energy $\epsilon_h= \gamma - \omega/2$. As a side remark, we note that \textit{anomalous velocity} terms $\delta \v v \sim \boldsymbol{\Omega}_\xi \times \v E$ are unimportant for the photocarriers, since their nonequilibrium distribution function Eq.~\eqref{eq:deltaf} is quadratic in electric field.

The final steady state current can be written as
\begin{subequations}
\begin{eqnarray}
\v j &=& - {\frac{e^3}{12 \pi}} \Bigg \lbrace \tau_{\rm CPGE}   [-i (\boldsymbol{\mathcal E}_+ \times\boldsymbol{\mathcal E}_-)] \notag \\
&& +\frac{u}{v} \Bigg [\tau_{\rm skew}   [-i (\boldsymbol{\mathcal E}_+ \times\boldsymbol{\mathcal E}_-) \times \hat t] \notag \\
&& - \tau_{\hat t}\hat t (\boldsymbol{\mathcal E}_+ \cdot \boldsymbol{\mathcal E}_-) + \tau_{\perp} \text{Re}[\boldsymbol{\mathcal E}_+ \otimes \boldsymbol{\mathcal E}_- ] \hat t   \Bigg ]\Bigg \rbrace. \label{eq:SteadyStatePGE}
\end{eqnarray}
We present the microscopic values of the effective scattering rates $\tau_{\rm CPGE},\tau_{\rm skew},\tau_{\hat t},\tau_{\perp}$ in Eqs.~\eqref{eq:TauCPGENOScatt}-\eqref{eq:TauTensorWScatt}, below. In Appendix~\ref{app:Relaxation}, we generalize the calculation presented here to a finite intervalley scattering described by the parameter $\kappa = \vert \mathcal{V}_{\bm{ b}} /\mathcal{V}_0 \vert ^2$. The latter interpolates between long-range impurities ($\kappa = 0$, no intervalley scattering) and short range impurities ($\kappa =1$, strong intervalley scattering). 
The relaxation times in the limit of weak intervalley scattering $\kappa \rightarrow 0$ behave as
\begin{align}
&\tau_{\rm CPGE} \simeq \frac{1}{2}\sum_{\epsilon = \epsilon_e, \epsilon_h} 1/\Gamma_+ , \label{eq:TauCPGENOScatt}\\
&\tau_{\rm skew} \simeq \frac{1}{2}\sum_{\epsilon = \epsilon_e, \epsilon_h}  \Gamma_{\text{sk},+}/\Gamma_+^2 , \\
&\tau_{\hat t} \simeq -\frac{1}{2}\sum_{\epsilon = \epsilon_e, \epsilon_h} \frac{2}{3 \Gamma_- \kappa } \text{sgn}\left (\epsilon-{\gamma}\right) , \label{eq:SteadyStatePGE_taut}\\
&\tau_{\perp } \simeq-\frac{\kappa}{2}\sum_{\epsilon = \epsilon_e, \epsilon_h} \left [\frac{2 \Gamma_-}{5\Gamma_+^2}+\frac{2}{5 \Gamma_-}\right ]\text{sgn}\left (\epsilon-{\gamma}\right).
\end{align}
In contrast, for $\kappa \rightarrow 1$ we obtain
\begin{align}
&\tau_{\rm CPGE}\simeq \sum_{\epsilon = \epsilon_e, \epsilon_h} \frac{1}{3[\Gamma_- + \Gamma_+]} ,\\
&\tau_{\rm skew}\simeq \sum_{\epsilon = \epsilon_e, \epsilon_h} (1-\kappa)\frac{2(\Gamma_{\text{sk},+} \Gamma_- + \Gamma_{\text{sk},-}\Gamma_+)}{9(\Gamma_+ + \Gamma_-)^2} , \\
&\tau_{\hat t} \simeq -\sum_{\epsilon = \epsilon_e, \epsilon_h} \frac{9 \Gamma_-+5 \Gamma_+}{15 \Gamma_- (\Gamma_++\Gamma_-)}\text{sgn}\left (\epsilon-{\gamma}\right),\\
&\tau_{\perp } \simeq-\sum_{\epsilon = \epsilon_e, \epsilon_h}\frac{2}{15 (\Gamma_+ + \Gamma_-)}\text{sgn}\left (\epsilon-{\gamma}\right). \label{eq:TauTensorWScatt}
\end{align}
\label{eq:SteadyStatePGE_ALL}
\end{subequations}
General formulae which interpolate between these two limits are presented in Eq.~\eqref{eq:FullRelaxTimes} of Appendix~\ref{app:Relaxation} and are plotted in Fig.~\ref{fig:PlotOfTaus}.
Note that all relaxation rates $\Gamma_\zeta (\epsilon)$ and $\Gamma_{\rm sk, \zeta} (\epsilon)$ introduced in these formulae are implicitly energy dependent. However, in the photoactive cone, the particle-hole symmetry about the Weyl node implies for both photoelectrons and photoholes equal scattering rates $\Gamma_+ (\epsilon_e) = \Gamma_+ (\epsilon_h)$, $\Gamma_{\rm sk, +} (\epsilon_e) = \Gamma_{\rm sk, +}(\epsilon_h) $. The rate $\tau_{\hat t}$ is not defined in a noninteracting model without intervalley scattering as it is determined by the relaxation time of the isotropic part in Eq.~\eqref{eq:Sinj}. This can only be achieved by inelastic scattering or by means of intervalley scattering. We discuss this issue and other physical implication of the result presented in Eqs.~\eqref{eq:SteadyStatePGE_ALL}, Eq.~\eqref{eq:FullRelaxTimes} and Fig.~\ref{fig:PlotOfTaus} in Sec.~\ref{sec:Discussion}.

In Appendix~\ref{app:Relaxation} we also present a derivation of the time evolution after a sudden illumination in the limiting case $u/v = 0$. Technically, this amounts to adding the solution of the homogenoeous Boltzmann equation (Liouville equation), Eq.~\eqref{eq:Boltzmann}, to the particular steady state solution, Eq.~\eqref{eq:deltaf}. In terms of the final result Eq.~\eqref{eq:SteadyStatePGE} this amounts to the following replacement $\tau_{\rm CPGE} \rightarrow T(t)$ with 
\begin{equation}
T(t) \simeq \!\! \sum_{\epsilon = \epsilon_e, \epsilon_h}\!\!\begin{cases} \frac{1}{2\Gamma_+}\left[1- e^{- \Gamma_+ t}\right], & \kappa \rightarrow 0, \\
 \frac{1}{3(\Gamma_++\Gamma_-)}\left[1- e^{- 3 t (\Gamma_++\Gamma_-)/2}\right], & \kappa \rightarrow 1,
\end{cases} \label{eq:TimeEvolution}
\end{equation}
where we assumed the light to be switched on at time $t = 0$. For a more general formula of $T(t)$ which interpolates between the two limits $\kappa = 0$ and $\kappa = 1$, we refer the reader to Eq.~\eqref{eq:TdepGeneral} of Appendix \ref{app:Relaxation} as well as to Fig.~\ref{fig:TimeEvolution}. Again, the discussion of this result is relegated to Sec.~\ref{sec:Discussion}. 

Finally, the Appendix \ref{app:intraband} contains a calculation of the intraband rectified response due to particle hole excitations near the Fermi surface. There we show, that this contribution vanishes in the absence of tilt and intervalley scattering and that it is generically suppressed by the parameter $\Gamma_+/\omega \ll 1$.

\begin{figure}
\includegraphics[scale=.65]{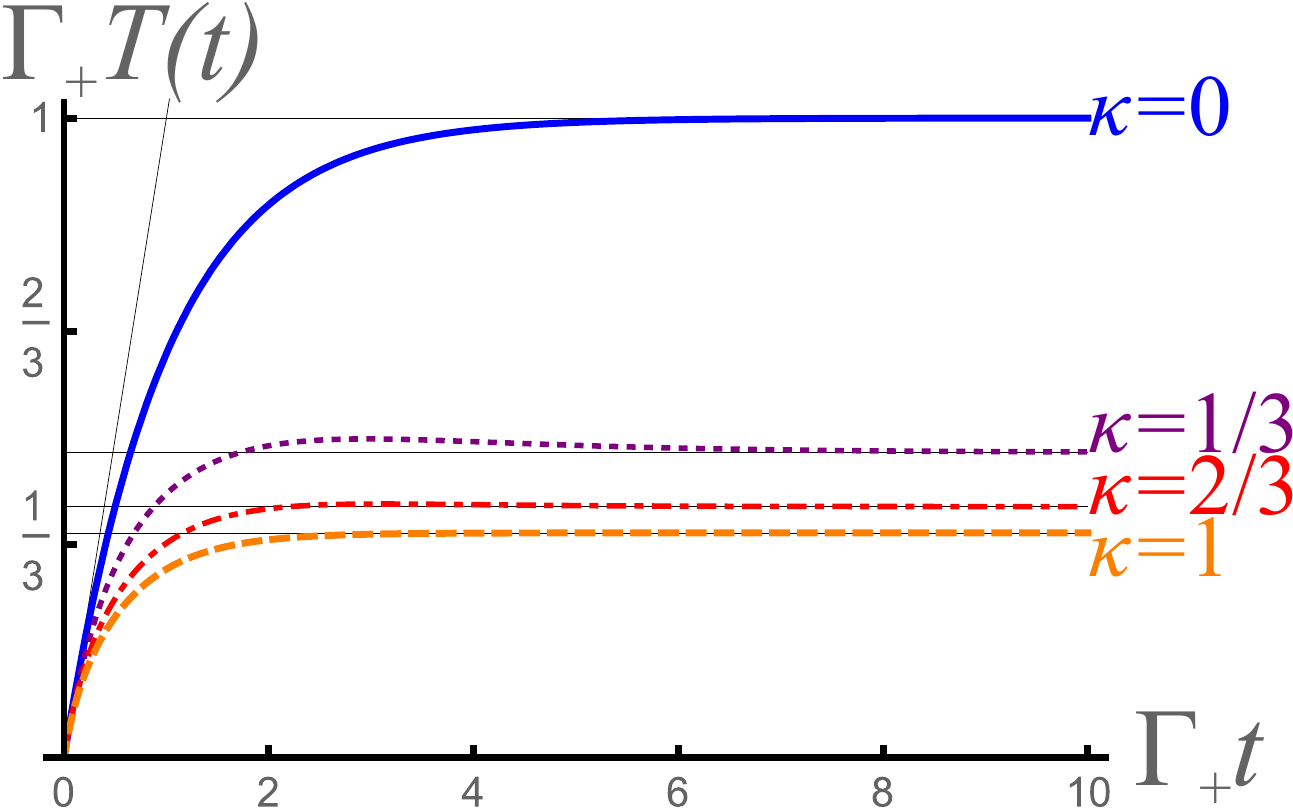} 
\caption{[Color online] Time evolution of the circular photogalvanic effect after a sudden illumination at time $t = 0$ for  $\kappa = 0,\, 1/3, \, 2/3,\, 1$ from top to bottom. In this plot $\omega = 3 \gamma$ was chosen, i.e. the case represented in Fig.~\ref{fig:Opttransition}.}
\label{fig:TimeEvolution}
\end{figure}

\section{Discussion}
\label{sec:Discussion}

In this section we analyze and discuss the physical content of the major results of this work, Eqs.~\eqref{eq:SteadyStatePGE_ALL} and \eqref{eq:TimeEvolution} as well as Figs.~\ref{fig:PlotOfTaus} and \ref{fig:TimeEvolution}. 

\subsection{Limit $u/v = 0$}
We begin the discussion considering the limit of absent tilt $u/v =0$. In this case, the dc photocurrent is proportional to $-i (\boldsymbol{\mathcal E}_+ \times \boldsymbol{\mathcal E}_-)$ and is thus crucially dependent on \textit{circularly} polarized light. This is the injection mechanism: vertical transitions inject photocarriers moving predominantly in direction $-i (\boldsymbol{\mathcal E}_+ \times \boldsymbol{\mathcal E}_-)$, see Eq.~\eqref{eq:Sinj}. This leads to the linearly increasing time-dependent response, Fig.~\ref{fig:TimeEvolution} at small $t \ll 1/\Gamma_+$. The slope is topologically quantized [\onlinecite{deJuanMoore2017}] in the present case of a single photoactive Weyl node. Disorder exponentially relaxes the linear increase and a steady state is formed, see Fig.~\ref{fig:TimeEvolution} at large $t \gg 1/\Gamma_+$. It is worthwhile to notice that $1/\Gamma_+$ is the ``transport'' mean free time, which is a factor of 3 larger than the ``quantum'' mean free time responsible for the level broadening. While these features are generic to any disorder model, intervalley scattering has the following three additional effects. First, it increases the scattering rate by increasing the density of available final states. This implies faster relaxation and a smaller steady state value. Second, since the relaxation rates at energies $\epsilon_e$ (photoelectrons) and $\epsilon_h$ (photoholes) are different in the $\zeta = -1$ node, intervalley scattering introduces two different relaxation rates for the circular PGE. Third, for intermediate $\kappa$ and provided $\Gamma_- < \Gamma_+$ intervalley scattering can lead to a peculiar nonmonotonic time dependence, see Eq.~\eqref{eq:TdepGeneral} and the $\kappa = 1/3$ and $\kappa = 2/3$ curves in Fig.~\ref{fig:TimeEvolution}. For the parameters chosen in our plot the nonmonotonicity stems from the photoholes living at energy $\epsilon_h$. Those scatter from the photoactive $\zeta = +1$ node to the $\zeta = -1$ node where the current is opposite and the decay much slower in view of a smaller DOS. 

\subsection{Finite $u/v$}

\begin{figure}
\includegraphics[scale=.55]{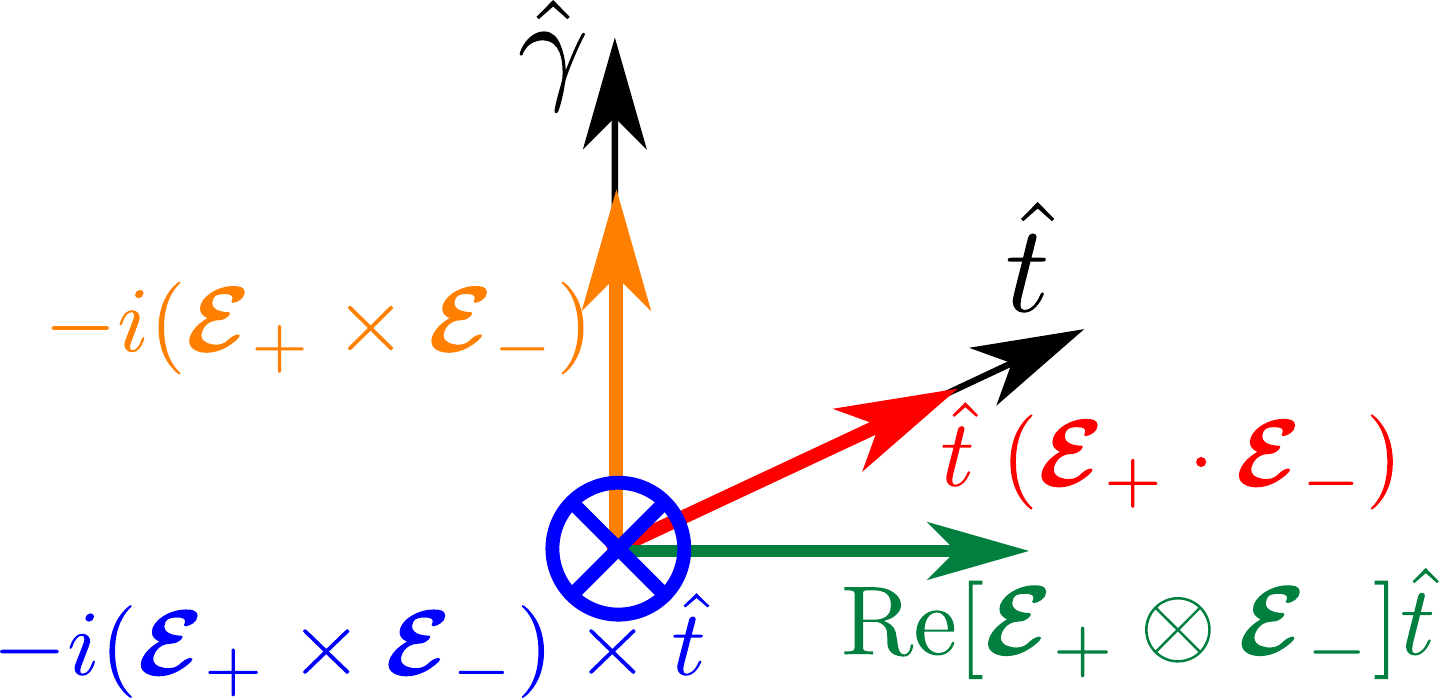} 
\caption{[Color online] Directions of photo current according to the final result, Eq.~\eqref{eq:SteadyStatePGE} (same color coding as Fig.~\ref{fig:PlotOfTaus}). Here, $\hat \gamma$ is the direction of incident light. The orange $-i (\boldsymbol{\mathcal{E}}_+ \times \boldsymbol{\mathcal{E}}_-)$ term is the ``quantized'' response for circularly polarized light [\onlinecite{deJuanMoore2017}]. The blue vector $-i(\boldsymbol{\mathcal{E}}_+ \times \boldsymbol{\mathcal{E}}_-) \times \hat t$ is directed into the plane spanned by the directions of light propagation and tilt $\hat t$. Just as the green $ \text{Re}[\boldsymbol{\mathcal E}_+\otimes\boldsymbol{\mathcal E}_-]\hat t $ and red $ \hat t (\boldsymbol{\mathcal{E}}_+ \cdot \boldsymbol{\mathcal{E}}_-)$ vectors it is a consequence of finite tilt. In the absence of intervalley scattering, the red contribution proportional to $\hat t$ can not be relaxed by impurities and may therefore be dominant, see Fig.~\ref{fig:PlotOfTaus}.}
\label{fig:DirectionsofPGE}
\end{figure}

We now turn to the steady state solution at finite $u/v$. The finite tilt allows for the presence of additional contributions to the current, see Figs.~\ref{fig:PlotOfTaus} and \ref{fig:DirectionsofPGE}. 

First, skew scattering leads to a term $-i (\boldsymbol{\mathcal E}_+ \times \boldsymbol{\mathcal E}_-) \times \hat t$. Just as the circular PGE, it is only present for circular polarization of light. As we already mentioned, the momentum of excited photo electrons predominantly points in direction $-i (\boldsymbol{\mathcal E}_+ \times \boldsymbol{\mathcal E}_-)$. The time-reversal symmetry breaking tilt introduces a finite anisotropy leading to a preferred direction of momentum relaxation. The resultant imbalance between electrons moving towards $-i (\boldsymbol{\mathcal E}_+ \times \boldsymbol{\mathcal E}_-) \times \hat t$ as compared the opposing direction generates the skew scattering term proportional to $\tau_{\rm skew}$. It is in accordance with intuitive expectation that opposite tilts in opposite nodes imply antagonistic skew scattering contributions from opposite Weyl nodes. It is however surprising to find, that the $\mathcal O(u/v)$ skew scattering contribution exactly vanishes at strongest intervalley scattering $\kappa = 1$, see the blue dot-dashed curve in Fig.~\ref{fig:PlotOfTaus}. 

Apart from skew scattering, the finite tilt in the spectrum also introduces terms which do not rely on circular polarization of light and which may be present also for linearly polarized light. We first focus on the term proportional to $\hat t (\boldsymbol{\mathcal E}_+ \cdot \boldsymbol{\mathcal E}_-)$. This term stems from the $(u/v) \hat t$ term in the definition of the current, Eq.~\eqref{eq:CurrentDef}. Therefore, it stems from the scalar (isotropic) part of the distribution function of photocarriers, or differently said, from the scalar term proportional to $\mathbf 1$ in Eq.~\eqref{eq:Sinj}. As we already mentioned, without intervalley scattering impurities cannot relax such a term. This is why the photocurrent in direction $\hat t$, represented by a red arrow in Fig.~\ref{fig:DirectionsofPGE}, is much stronger than all other contributions. The presence of intervalley scattering $\kappa >0$ introduces a finite $\tau_{\hat t}$, which however diverges as $1/\kappa$ for small $\kappa$, see Eq.~\eqref{eq:SteadyStatePGE_taut}. This is represented in the red dashed curve in Fig.~\ref{fig:PlotOfTaus} which, for the sake of a better presentation, had to be downscaled by an extra factor of $20$. In the chosen parameter regime, $\tau_{\hat t}$ is particularly large, because $\Gamma_-$ is relatively small at $\epsilon_h$. Even though inelastic scattering is beyond the scope of the present paper, we mention that in practice, the steady state value of the current in $\hat t$ direction is determined by the smaller of $\tau_{\rm inel}$ (inelastic scattering time) and $\tau_{\hat t}$.

Finally, the steady state result, Eq.~\eqref{eq:SteadyStatePGE} contains a term proportional to $\text{Re}[\boldsymbol{\mathcal E}_+\otimes\boldsymbol{\mathcal E}_-]\hat t$. This term stems from the tensor contribution to the photocarrier excitation rate, i.e. the $\hat p \otimes \hat p$ term in Eq.~\eqref{eq:Sinj} and is absent in the absence of intervalley scattering.

\section{Summary and Outlook}
\label{sec:outlook}

In summary, we have derived and analyzed the disorder induced relaxation of the photogalvanic effect (PGE) in a simple model for a Weyl semimetal. We took into account intra and intervalley scattering as well as leading order corrections due to finite tilt of the Weyl spectrum. 

The major findings, which are pictorially summarized in Figs.~\ref{fig:PlotOfTaus}-\ref{fig:DirectionsofPGE}, are that (i) intervalley scattering can lead to nonmonotonic time dependence of the circular PGE proportional to $-i (\boldsymbol{\mathcal E}_+ \times \boldsymbol{\mathcal E}_-)$; (ii) the finite tilt introduces additional current components proportional to (a) $-i (\boldsymbol{\mathcal E}_+ \times \boldsymbol{\mathcal E}_-)\times \hat t$ due to skew scattering, (b) $\hat t (\boldsymbol{\mathcal E}_+ \cdot \boldsymbol{\mathcal E}_-)$, and (c) $\text{Re}[\boldsymbol{\mathcal E}_+\otimes\boldsymbol{\mathcal E}_-]\hat t $. As we discussed, the contribution (ii b) in the direction of the tilt vector $\hat t $ is particularly inert to elastic scattering as it stems from isotropic generation of photo carriers and may thus be dominant. Contributions (i) and (ii a) involving $-i (\boldsymbol{\mathcal E}_+ \times \boldsymbol{\mathcal E}_-)$ change sign when the chirality of the photoactive Weyl node is reversed.

The observation of decisive qualitative and quantitative consequences of a weak spectral tilt corroborates the findings of earlier studies on different observables such as the linear conductivity tensor [\onlinecite{TrescherBergholtz2015,Carbotte2016,SteinerPesin2017}], the polarization function [\onlinecite{DetassisGrubinskas2017}] as well as transport through Weyl tunnel junctions [\onlinecite{YesilyurtJalil2017}]. Further implications on characteristic features of Weyl fermions, e.g. the natural optical activity [\onlinecite{MaPesin2015}], will be the subject of a separate studies [\onlinecite{RouPesin2017}].

We conclude our paper with an outlook on the experimental relevance of our findings. First of all, we comment on the minimal two Weyl node model that we are considering. Such materials are not discovered so far, but they were suggested theoretically in heterostructures [\onlinecite{BurkovBalents2011}] and alloys [\onlinecite{BulmashQi2014}] of Chern insulator materials as well as, most recently, in certain magnetic Heusler compounds [\onlinecite{WangBernevig2016}]. At the same time our simplified model may be applied to present day materials when pairs of Weyl nodes are well separated in momentum space.

Concerning optical experiments, we are aware of only two published results on the PGE in Weyl semimetals. The work of Ref. [\onlinecite{WuOrenstein2016}] is devoted to the second harmonic generation, i.e. a different observable as compared to the focus of this paper. In the second experiment [\onlinecite{Jarillo-Herrero-Gedik2017}] the chirality of the Weyl fermions, which is given by the sign of the Berry monopole charge, was inferred from the photocurrent response to mid-infrared light. Two key points are important: (i) Experiment reveals a current component perpendicular to the direction of light which strongly depends on the polarization (linear, circular positive, circular negative) and which is claimed to be proportional to the chirality of the photoactive nodes. As we show, skew scattering also induces a term of the very same tensor structure. (ii) Polarization independent photocurrent is observed in a direction perpendicular to both light and to the polarization dependent current. As we show, such terms may also be induced by the tilt. In addition, unpublished THz spectroscopy data on the dc PGE in TaAs presented at the APS march meeting 2017 [\onlinecite{Patankar2017}] indicates that the signals of radiated electric field in direction perpendicular or parallel to the polar axis show very different behavior and that for the latter case the signal is nearly independent of the polarization of incident light and is relatively long lived. The preliminary interpretation focused on anisotropic scattering and is thus related to the skew scattering mechanism investigated here. Another route for interpretation could be contributions similar to the terms $\hat t (\boldsymbol{\mathcal E}_+ \cdot \boldsymbol{\mathcal E}_-)$ considered in this paper. We repeat that these are indeed long lived and independent on the polarization of incident light. At the same time, we note that the tilt of the Weyl fermions in TaAs is substantial ($v/u>2$) and, therefore, our perturbative calculations should not be expected to quantitatively describe those experiments since our time-reversal breaking two node toy model has limited resemblance with the time reversal conserving 24 Weyl node material TaAs. However, we are convinced that the present results provide a proof of principle for various types of photocurrent contributions which will trigger future investigations.

\section{Acknowledgments}

We acknowledge useful discussions with A.V. Andreev, P.-Y.~Chang, P.~Coleman, L.~Golub, A.~Grushin, Y.~Komijani, J.~Moore, A.~A.~Zyuzin, and, in particular, with S.~Patankar and L.~Wu, who shared insights from ongoing experimental efforts.

The work of E.J.K., H.-Y.X. and A.L. at the University of Wisconsin-Madison was financially supported in part by NSF Grant No. DMR-1606517 and by the Office of the Vice Chancellor for Research and Graduate Education with funding from the Wisconsin Alumni Research Foundation. The work of D.A.P. at the University of Utah was supported by the NSF Grant No. DMR-1409089. At the latest stage of the project (when writing the manuscript), work by E.J.K. was carried out at Rutgers University, where financial support was provided by DOE, Basic Energy Sciences grant DE-FG02-99ER45790.

\appendix

\section{Model}
\label{app:model}

It is useful to perform a gauge transformation of Eq.~\eqref{eq:H}, such that 
\begin{subequations}
\begin{align}
H_{\rm kin} &={\left (\begin{array}{cc}
u \hat t \v p + v \v p  \cdot \boldsymbol \sigma + \gamma & 0 \\ 
0 & - u \hat t \v p -v \v p \cdot \boldsymbol \sigma - \gamma
\end{array} \right )_\kappa},  \\
H_{\rm dis} &={\left (\begin{array}{cc}
V(\v x) & V(\v x) e^{2 i \v b \v x} \\ 
V(\v x) e^{-2 i \v b \v x} & V(\v x)
\end{array} \right )_\kappa}.
\end{align}
\end{subequations}
The potential term $\Phi(\v x, t)$ remains unaffected from the gauge transformation.
The eigenstates of the clean Hamiltonian are characterized by the quantum number (multiindex) $l = (\zeta, \xi, \v p)$, with $\zeta = \pm 1$, $\xi = \pm 1$ and $\v p \in \mathbb R^3$: 

\begin{equation}
\psi_l(\v x) = \braket{\v x \vert l} = e^{i \v p \v x} \ket{ u_{\xi, \v p}} \hat e_\zeta,
\end{equation}
where $\hat e_+ = (1,0)_\kappa$ and $\hat e_- = (0,1)_\kappa$ and
\begin{equation}
\ket{ u_{\xi, \v p}} = \frac{1}{\sqrt{2 (1 + \xi \hat p_3)}} \left (\begin{array}{c}
\xi + \hat p_3 \\ 
\hat p_1 + i \hat p_2
\end{array} \right )_\sigma .
\end{equation}

The unperturbed eigenenergies associated to the eigenstates $\ket{l}$ are 
\begin{equation}
\epsilon_l = \zeta (u \hat t \cdot \v p + \xi v p + \gamma). \label{eq:Energy}
\end{equation} 

The following relations will be useful  ($\hat p_1 + i \hat p_2 = \sqrt{1 - \hat p_3^2}e^{i \phi}$):
\begin{subequations}
\begin{eqnarray}
\braket{u_{\xi, \v p} \vert u_{\xi', \v p'}} &=& \Big[ \sqrt{(1- \xi \hat p_3)(1- \xi '\hat p_3')} e^{i(\phi' - \phi)}\notag \\
&+&\xi \xi' \sqrt{(1+ \xi \hat p_3)(1+ \xi '\hat p_3')}\Big ]/{2}, \\
\vert \braket{u_{\xi, \v p} \vert u_{\xi', \v p'}} \vert^2& =& \frac{1 + \xi \xi' \hat p \cdot \hat p'}{2},\\
\ket{u_{\xi, \v p}} \bra{u_{\xi, \v p}} &=& \frac{\mathbf 1 + \xi \hat p \cdot \boldsymbol \sigma }{2}, \label{eq:ketbra}\\
\boldsymbol \Omega_{\xi}(\v p) &=&  -\xi  \frac{\hat p}{2p^2}.
\end{eqnarray}
\end{subequations}

\subsection{Deformed coordinate system}

In view of the tilt, the Fermi surfaces are ellipsoidal. Here we introduce a coordinate transformation which deforms the ellipse into a circle at the expense of certain Jacobians.

We first use that at a given energy $\epsilon$ the equienergy surface of a band characterized by $(\xi,\zeta)$ is an ellipse in the $p_t - p_\perp$ plane ($p_t = \hat t \cdot \v p$, $p_\perp = \sqrt{p^2 - p_t^2}$) with center at $(p_t, p_\perp) = \xi (\zeta \epsilon - \gamma)( \frac{-\xi u}{v^2 - u^2},0)$. The radius in $p_t$ ($p_\perp$) direction is $\xi (\zeta \epsilon - \gamma) \frac{v}{v^2- u^2}$ ($\xi (\zeta \epsilon - \gamma) \frac{1}{\sqrt{v^2- u^2}}$).

It is useful to choose the following energy dependent parametrization of the three vector $\v p$ . 
\begin{equation}
\v p = \frac{\xi (\zeta \epsilon - \gamma)}{v} \left \lbrace
\v k  - \xi \hat t \frac{u}{v} \right \rbrace.
\end{equation}

The integration measure transforms as 
\begin{equation}
(dp) = \xi (\zeta \epsilon - \gamma) \nu_{\xi \zeta}(\epsilon) \frac{d^3 k}{4 \pi}, 
\end{equation}
and, since a delta function in energy space can be written as
\begin{equation}
\delta(\epsilon_l - \epsilon) = \frac{1 - \xi k_t u/v}{\xi (\zeta \epsilon - \gamma)} \delta (\vert \v k \vert - 1)
\end{equation}
we can write
\begin{equation}
\int (dp) \delta(\epsilon_l - \epsilon)  \dots = \nu_{\xi \zeta} (\epsilon) \left \langle (1 - \xi \frac{ u}{v} \hat k \cdot \hat t )\dots \right \rangle_{\hat k} .
\end{equation}
Note that, when $\v k$ is a unit vector, we can write
\begin{equation}
\hat p = \hat k + \xi \frac{u}{v} [ \hat k(\hat k \cdot \hat t )- \hat t].
\end{equation}

We will use the following parametrization of $f_{\xi, \zeta}(\v p)$
\begin{equation}
f_{\xi,\zeta} (\v p) = f_{\xi, \zeta} (\epsilon, \hat k),
\end{equation}
where the transformation $\v p \rightarrow (\epsilon,\hat k)$ depends on the cone $\xi, \zeta$ under consideration. We repeat that the band index $\xi$ is uniquely determined by $\epsilon$ and $\zeta$.

\section{Elastic scattering rates}
\label{app:Rates}

The scattering probability for scattering from the state $l$ to the state $l'$ is
\begin{equation}
w_{l'l} = 2 \pi \vert T_{l'l} \vert^2 \delta(\epsilon_l - \epsilon_l'), \label{eq:ScattProb}
\end{equation}
where
\begin{equation}
T_{l'l} = \braket{l'\vert \underbrace{\hat H_{\rm dis}+\hat H_{\rm dis} G^R(\epsilon_l) \hat H_{\rm dis}}_{= \hat T^R} \vert l}
\end{equation}
are the matrix elements of the retarded transition matrix. We next present restrictions on the scattering probability imposed by the symmetries discussed in the main text. 

\subsection{Symmetry considerations}
 
 As we mentioned in the main text, our model is not invariant under any of time reversal, rotation or inversion symmetries. We find
\begin{subequations}
\begin{eqnarray}
\hat T H_0  \hat T&=& \left .  H_0\right \vert_{\v b \rightarrow - \v b; \hat t \rightarrow - \hat t} ,\\
\hat I H_0\hat I  &=& \left . H_0\right \vert_{{\gamma \rightarrow - \gamma ;  V( \v x) \rightarrow V(- \v x)}} ,\\
\hat R_U^\dagger H_0 \hat R_U &=& \left . H_0 \right \vert_{{\v b \rightarrow \underline R_U^T \v b ; \hat t \rightarrow \underline R_U \hat t;  V( \v x) \rightarrow V(\underline R_U \v x)}}.
\end{eqnarray}
\end{subequations}
We used the representations $\hat T = \sigma_y K$ (K is the complex conjugation); $\hat I = \kappa_1 \mathcal I$ ($\mathcal I: \v x \rightarrow - \v x$) and $\hat R_U = U \mathcal R_{\underline R_U^T}$ ($\mathcal R_{\underline R_U} : \v x \rightarrow \underline R_U \v x$ and $U \boldsymbol \sigma U^\dagger = \underline R_U\boldsymbol \sigma$ is the $\text{SU}(2)$-$\text{SO}(3)$ homomorphism). 

The symmetry operations on the wave function are as follows
\begin{subequations}
\begin{eqnarray}
\hat T \psi_{\zeta, \xi, \v p}(\v x) &=& - i e^{-i \phi} \xi \psi_{\zeta, \xi, - \v p} (\v x) ,\\
\hat I \psi_{\zeta, \xi, \v p}(\v x) &=& - \psi_{- \zeta, - \xi, - \v p} ( \v x) ,\\
\hat I \hat T \psi_{\zeta, \xi, \v p } (\v p)&=& i e^{- i \phi} \xi \psi_{-\zeta, - \xi, \v p} (\v x).
\end{eqnarray}
\label{eq:TandIactonpsi}
\end{subequations}
Here, again we used $\hat p_1 + i \hat p_2 = (1 - \hat p_3^2)e^{i \phi}$. For the rotational symmetry it is more useful to investigate its action on a projector onto an eigenstate
\begin{equation}
\hat R_U [\psi_l(\v x) \psi_l(\v x')^\dagger] \hat R_U^\dagger=\left . [ \psi_l(\v x) \psi_l(\v x')^\dagger]\right \vert_{\v p \rightarrow \underline R_U^T \v p} . \label{eq:RotOnProj}
\end{equation}

We exploit those transformations in the analysis of the scattering probabilities in the limiting case $u/v = 0$. For the present model, rotational symmetry plays a major role and we find that, for $u = 0$ and after average over disorder configuration,
\begin{equation}
\vert T_{l'l} \vert^2  = \vert T_{l'l} \vert^2_{\v p' \rightarrow \underline R_{U}^T \v p'; \v p \rightarrow \underline R_{U}^T \v p; \v b\rightarrow \underline R_{U}^T \v b} ,
\end{equation}
which implies that $\overline{w_{l'l}}$ is a scalar or pseudo scalar under transformations and can be expanded by the basis functions $1, \hat p \cdot \hat p', \hat p \cdot \v b, \hat p' \cdot \v b, (\hat p \times \hat p') \cdot \v b$. We can use the fact that after disorder average, the average translational inveriance implies that $\v b$ enters only in the form of $\vert \mathcal{V}_{\bm b} \vert^2$ under the assumption of sufficiently short ranged impurities. Therefore, rotational invariance implies that 
\begin{equation}
{{w_{l'l}} = [f_{\zeta, \xi; \zeta', \xi'} + g_{\zeta, \xi; \zeta', \xi'} \hat p \cdot \hat p']\delta(\epsilon_l- \epsilon_{l'}).}
\end{equation}
In particular, $w_{l'l}$ is symmetric under the exchange of momenta.

We can further use the implication of TR symmetry and find
\begin{equation}
T_{l'l} = \left . T_{ll'}\right \vert_{\v p \rightarrow - \v p, \v p' \rightarrow - \v p', \v b \rightarrow -\v b}. \label{eq:TRsymm}
\end{equation}
so that
\begin{equation}
{{w_{l'l}} = [f_{\zeta', \xi';\zeta, \xi} + g_{\zeta', \xi';\zeta, \xi} \hat p \cdot \hat p'}]\delta(\epsilon_l- \epsilon_{l'})
\end{equation}
and thus the absence of skew scattering: ${w_{l'l}} = {w_{ll'} }$ in the limit $u = 0$.

\subsection{Scattering probabilities}
\subsubsection{Born approximation}
The scattering probability in Born approximation can be readily evaluated from Eq.~\eqref{eq:ScattProb}
\begin{equation}
w_{l'l}^{(2s)} = 2\pi \delta (\epsilon_l - \epsilon_{l'}) n_{\rm imp} \vert \mathcal{V}_{(\zeta - \zeta') \bm b} \vert ^2 \frac{1 + \xi \xi' \hat p \cdot \hat p'}{2}.
\end{equation}

\subsubsection{Skew scattering}

For the skew scattering we expand $T_{l'l} \simeq \braket{l' \vert \hat H_{\rm dis} + \hat H_{\rm dis} G_0^R \hat H_{\rm dis} \vert l}$ and obtain [\onlinecite{NagaosaOng2010}]
\begin{eqnarray}
w_{l'l}^{(3a)} &=& (2\pi)^2 \delta(\epsilon_l - \epsilon_{l'}) \sum \hspace{-.4cm} \int_{l''} \delta(\epsilon_l -\epsilon_{l''}) \notag \\
&&\times \text{Im}[\hat H_{\text{dis}, ll'} \hat H_{\text{dis}, l'l''} \hat H_{\text{dis}, l'' l}].
\end{eqnarray}
 For our specific model this leads to
\begin{eqnarray}
&&w_{l'l}^{(3a)} = (2 \pi)^2 \delta (\epsilon_l - \epsilon_{l'}) \sum_{\xi'' \zeta''} \int_{\v p''} \delta(\epsilon_{l''} - \epsilon_l) \notag \\
&&\times  n_{\rm imp} \mathcal{V}_{(\zeta - \zeta') \bm b}\mathcal{V}_{(\zeta' - \zeta'') \bm b}
\mathcal{V}_{(\zeta'' - \zeta) \bm b}  \notag \\
&& \times \text{Im}[\braket{u_{\xi \v p} \vert u_{\xi' \v p'}} \braket{u_{\xi' \v p'} \vert u_{\xi'' \v p''}} \braket{u_{\xi'' \v p''} \vert u_{\xi \v p}}].
\end{eqnarray}
We use that 
\begin{equation}
\int_{\v p''}\delta(\epsilon_{l''} - \epsilon_l) \ket{u_{\xi'' \v p''}} \bra{u_{\xi'' \v p''}} =  \frac{1}{2} \nu_{\xi'' \zeta ''}(\epsilon_l) \left (1 - \frac{u}{v} \hat t \cdot \boldsymbol{\sigma}\right ),
\end{equation}
so that we find
\begin{eqnarray}
&&w_{l'l}^{(3a)} = (2 \pi) \delta (\epsilon_l - \epsilon_{l'}) \sum_{\xi'' \zeta''} \left (-\frac{\xi \xi' \pi u}{2 v }\right ) \nu_{\xi'', \zeta''} (\epsilon_l) \notag \\
&&\hat t \cdot (\hat p \times \hat p') n_{\rm imp} \mathcal{V}_{(\zeta - \zeta') \bm b}\mathcal{V}_{(\zeta' - \zeta'') \bm b}
\mathcal{V}_{(\zeta'' - \zeta) \bm b}. 
\end{eqnarray}
This concludes the derivation of Eq.~\eqref{eq:SkewProb} of the main text.

\subsection{Scattering rates}

It will be useful to expand the distribution function for the band $(\xi, \zeta)$ by means of representations of the rotation group. We write
\begin{eqnarray}
f_{\xi, \zeta} (\epsilon, \hat k) &=& f^{(sc)}_{\xi, \zeta} (\epsilon) + \hat k_i  f_{\xi, \zeta}^{i} (\epsilon) \notag \\
&&+ \hat k_i \hat k_j f_{\xi, \zeta}^{ij}(\epsilon)+ \hat k_i \hat k_j \hat k_k f_{\xi, \zeta}^{ijk}(\epsilon).
\end{eqnarray}
Furthermore, the kinetic equation and the collision integral shall be written as a vector $\underline{f} = (f_{\zeta = +1}, f_{\zeta = -1})^T$ in valley space and denote it by an underbar. Note that, at a given energy $\epsilon$ there are two bands involved, one at each valley.

\subsubsection{Born approximation}
We consider the Born scattering from impurities introduce scattering rates $\Gamma_\zeta = 2 \pi n_{\rm imp} \vert \mathcal{V}_0 \vert^2 \nu_{\xi, \zeta} (\epsilon)$ ($\xi$ is determined by $\zeta$ and $\epsilon$ uniquely).

We can write the collision integral entering the equation for electrons in pocket $\xi, \zeta$
\begin{widetext}
\begin{equation}
St^{(2)}[\lbrace f \rbrace] \vert_{\xi, \zeta}  = - \sum_{\xi', \zeta' }3 \Gamma_{\xi', \zeta'}  \frac{\vert \mathcal{V}_{(\zeta - \zeta ') \bm b} \vert^2}{\vert \mathcal{V}_0 \vert^2} \Big \langle \frac{1 - \xi' \frac{u}{v} ( \hat k' \cdot \hat t) + \xi \xi' \hat p(\hat k' - \xi' \frac{u}{v} \hat t)}{2} \left (f_{\xi, \zeta} - f_{\xi', \zeta'}\right )\Big \rangle_{\hat k'}.
\end{equation}
The notation $St[\lbrace f \rbrace]$ reflects that in general the collision integral depends on the whole set of distribution functions, i.e. in the present case on the distribution functions of both nodes.  All energies are taken at given $\epsilon_{\xi, \zeta}$ which is either $\epsilon_e$ or $\epsilon_h$. We employ the valley space vector notation for the scalar part of the distribution function to $\mathcal O(u/v)$
\begin{eqnarray}
\underline{St}^{(2)}[\lbrace f^{(sc)} \rbrace]  &=& - \frac{3}{2} \left (1 - \hat k \cdot \hat t \underline \xi \frac{u}{v}\right )\underbrace{\left (\begin{array}{cc}
\Gamma_- \kappa & - \Gamma_- \kappa \\ 
- \Gamma_+ \kappa & \Gamma_+ \kappa
\end{array} \right )}_{=: \underline{\Gamma}_s} 
\underline{f}^{(sc)} .
\end{eqnarray}
Here, the matrix $\underline \xi = \text{diag}(\xi_+, \xi_-)$. As we shall see later, for the vector part we don't need $u/v$ corrections which contain an even power of vectors $\hat k$. Under this premise (reflected by the sign $\doteq$), only the nontilted contribution survives
\begin{eqnarray}
\underline{St}^{(2)}[\lbrace \hat k_i f^{i} \rbrace]  &\doteq& - \underbrace{\underline{\xi} \left (\begin{array}{cc}
\Gamma_+ + 3\Gamma_- \kappa /2& - \Gamma_- \kappa/2 \\ 
- \Gamma_+ \kappa /2& \Gamma_- + 3\Gamma_+\kappa/2 
\end{array} \right ) \underline{\xi}}_{=: \underline{\Gamma}_v} \hat k_i  \underline{f}^{i}.
\end{eqnarray}
Next, we turn our attention to the matrix distribution function. Here we obtain
\begin{equation}
\underline{St}^{(2)}[\lbrace \hat k_i \hat k_j f^{ij} \rbrace]  = - \frac{3}{2} \left (1 - \hat k \cdot \hat t \underline \xi \frac{u}{v}\right )\underbrace{\left (\begin{array}{cc}
\Gamma_++\Gamma_- \kappa &0 \\ 
0& \Gamma_-+\Gamma_+ \kappa 
\end{array} \right )}_{=: \underline{\Gamma}_t} \hat k_i \hat k_j \underline{f}^{ij} + \frac{1}{2} \left (1 - \hat k \cdot \hat t \underline \xi \frac{u}{v}\right )\underbrace{\left (\begin{array}{cc}
\Gamma_+ & \Gamma_- \kappa \\ 
 \Gamma_+ \kappa & \Gamma_- 
\end{array} \right )}_{=: \underline{\Gamma}_{tr}} \underline{f}^{ii} .
\end{equation}
The sum over repeated indices is to be understood. We will find later that the solution for the tensorial part of the distribution function $f^{ijk}$ is first order in $u/v$. Therefore, we only need the zeroth order in $u/v$ from the collision integral, i.e.
\begin{equation}
\underline{St}^{(2)}[\lbrace \hat k_i \hat k_j\hat k_k f^{ijk} \rbrace]  = - \frac{3}{2} \underline{\Gamma}_t 
\hat k_i \hat k_j\hat k_k \underline{f}^{ijk} + \frac{1}{10} \underline{\xi} \underline{\Gamma}_{tr} \underline{\xi}  \underline{f}^{ijk} [\hat k_i \delta_{jk} + 2 \times \text{cycl.}].
\end{equation}

\subsubsection{Skew scattering}
 For the skew scattering the leading order is $(u/v)^1$ and we only need to consider the contributions of the scalar part (vanishes), of the vector part (nonvanishing) and of the matrix part (vanishes) of the distribution function.
 
 After angular integration the skew-scattering collision integral for electrons in band $(\xi, \zeta)$ becomes 
 \begin{equation}
 St^{(3a)} \vert_{\xi, \zeta} = \frac{u}{v} \sum_{\substack{\xi', \zeta' \\ \xi'', \zeta ''}}\xi \xi'  \nu_{\xi' \zeta'}\nu_{\xi''\zeta''} \frac{\pi^2}{3} \hat k \cdot (\v f_{\xi', \zeta'} \times \hat t)  n_{\rm imp} \mathcal{V}_{(\zeta - \zeta ') \bm b}\mathcal{V}_{(\zeta' - \zeta '') \bm b}\mathcal{V}_{(\zeta'' - \zeta ) \bm b}.
 \end{equation}

We define the band dependent skew scattering rate $\Gamma_{\zeta}^{\rm (sk)} =  \frac{\pi^2}{3} n_{\rm imp} \mathcal{V}_0^3 \nu_{\xi \zeta}^2$. We further use that $\Gamma_+/\Gamma_- = \nu_+/\nu_-$ to reexpress the ratios of DOSs. Then 

\begin{equation}
\underline{St}^{(3a)}[\lbrace \hat k_i f^i \rbrace] =\frac{u}{v} \underbrace{\underline{\xi} \left (\begin{array}{cc}
\Gamma_+^{\rm (sk)} (1 + \kappa \frac{\Gamma_-}{\Gamma_+}) & \Gamma_-^{\rm (sk)} (1 +  \frac{\Gamma_+}{\Gamma_-})\kappa  \\ 
 \Gamma_+^{\rm (sk)} (1 +  \frac{\Gamma_-}{\Gamma_+})\kappa& \Gamma_-^{\rm (sk)} (1 + \kappa \frac{\Gamma_+}{\Gamma_-})
\end{array} \right )\underline{\xi} }_{=: \underline{\Gamma}^{\rm (sk)}} \epsilon_{ijk}\hat{k}_i \underline{f}^j\hat t_k .
\end{equation}
\end{widetext}
Note that $\Gamma_+^{\rm (sk)}{\Gamma_-}/{\Gamma_+} = \Gamma_-^{\rm (sk)}{\Gamma_+}/{\Gamma_-}$.

\section{Relaxation of photocarriers}
\label{app:Relaxation}

We use the notation $\zeta_0 = \text{sgn}(\epsilon_F)$ for photoactive node and we further write $\xi_0$ for the band index in that node at a given energy $\epsilon_e$ or $\epsilon_h$. 

\subsection{Injection term}

The injection term, i.e. the collision term in the Boltzmann equation generating photoelectrons/photoholes in band $(\xi_0, \zeta_0)$ in vector notation is given by the Fermi's golden-rule expression
\begin{eqnarray}
\underline{St}_{\rm inj} [\lbrace f \rbrace] &=& - \underline{\hat e}_{\zeta_0} \sum_{\alpha = \pm 1} \Big \lbrace 2\pi \delta (\alpha \omega + \epsilon_{\xi_0 \zeta_0}(\v p) - \epsilon_{-\xi_0, \zeta_0}(\v p)) \notag \\
&&  \mathcal A_{(\xi_0, \zeta_0), (-\xi_0, \zeta_0)}^i \mathcal A_{(-\xi_0, \zeta_0), (\xi_0, \zeta_0)}^{j}\notag \\
&&\left [ \frac{(e \mathcal E_\alpha^i)(e \mathcal E_{\alpha}^{j})e^{i \alpha 2 \omega t} + c.c.}{2 } +  (e \mathcal E_{\alpha}^i)(e \mathcal E_{- \alpha}^{j}) \right ] \notag \\
&&[f^{(eq.)}_{\xi_0, \zeta_0}( \v p) - f^{(eq.)}_{-\xi_0, \zeta_0}( \v p) ] \Big \rbrace .
\end{eqnarray}
For the product of transition matrix elements 
\begin{align}
&\mathcal A_{(\xi_0, \zeta_0), (-\xi_0, \zeta_0)}^i \mathcal A_{(-\xi_0, \zeta_0), (\xi_0, \zeta_0)}^{j} \notag \\
 &= \braket{\partial_{p_i} u_{\xi_0, \v p} \vert u_{-\xi_0, \v p}}\braket{ u_{-\xi_0, \v p} \vert \partial_{p_j} u_{\xi_0, \v p} }
\end{align}
we calculate symmetric (round brackets) and antisymmetric (square brackets) components separately
\begin{subequations}
\begin{align}
\mathcal A_{(\xi_0, \zeta_0), (-\xi_0, \zeta_0)}^{(i} \mathcal A_{(-\xi_0, \zeta_0), (\xi_0, \zeta_0)}^{j)} &=  \frac{(\mathbf 1 - \hat p \otimes \hat p)_{ij}}{4 p^2} ,\\
\mathcal A_{(\xi_0, \zeta_0), (-\xi_0, \zeta_0)}^{[i} \mathcal A_{(-\xi_0, \zeta_0), (\xi_0, \zeta_0)}^{j]}&= - \frac{i}{2} \epsilon_{ijk} \Omega_{\xi_0,k}.
\end{align}
\end{subequations}
This leads to Eq.~\eqref{eq:Sinj} of the main text. After the change of coordinates from $\hat p$ to $\hat k$ the injection term in vector notation can be written as
\begin{align}
\underline{St}_{\rm inj} &= \hat{\underline{e}}_{\zeta_0} \delta (\omega - 2 v p) \Bigg \lbrace I^{(0)} + \hat k_i I^{(0)}_i + \hat k_i \hat k_j I^{(0)}_{ij} \notag \\
&+ \frac{u}{v}\left [I^{(1)} + \hat k_i I^{(1)}_i + \hat k_i \hat k_j I^{(1)}_{ij} +\hat k_i \hat k_j \hat k_k I^{(1)}_{ijk}\right ]\Bigg \rbrace,
\end{align}
where we introduced (recall $\mathbf E(t) = \sum_\pm \boldsymbol{\mathcal E}_\pm e^{\pm i \omega t}$)
\begin{subequations}
\begin{eqnarray}
I^{(0)} &=& \frac{2\pi v^2}{\omega^2} \xi_0 \zeta_0 e^2 \left \lbrace \frac{\mathbf E(t)^2}{2}\right\rbrace ,\\
I^{(0)}_i &=& \frac{2\pi v^2}{\omega^2} \xi_0 \zeta_0 e^2 \left \lbrace - i \zeta_0 (e \boldsymbol{\mathcal E}_+ \times e\boldsymbol{\mathcal E}_-)_i\right\rbrace ,\\
I^{(0)}_{ij} &=& \frac{2\pi v^2}{\omega^2} \xi_0 \zeta_0 \left \lbrace -\frac{\mathbf E(t)_i \mathbf E(t)_j}{2}\right\rbrace ,
\end{eqnarray}
as well as
\begin{eqnarray}
I^{(1)} &=& - \xi_0 \hat t_i I_i^{(0)},\\
I^{(1)}_i &=&- \xi_0 \hat t_j (I_{ij}^{(0)}+I_{ji}^{(0)}) = - 2\xi_0 \hat t_j I_{ij}^{(0)}, \\
I^{(1)}_{ij} &=&  \xi_0 \hat t_j I_i^{(0)},\\
I^{(1)}_{ijk} &=&  \xi_0 \hat t_k (I_{ij}^{(0)}+I_{ji}^{(0)}) = 2 \xi_0 \hat t_k I_{ij}^{(0)}.
\end{eqnarray}
\end{subequations}
As we mentioned previously, we are only interested in the dc part of these expressions since $\omega/\Gamma_{\zeta_0} \gg 1$ implies that second harmonic generation is only weakly affected by disorder.

\subsection{Static solution to $\mathcal O[(u/v)^0]$}

We expand $\underline f = \underline f^{(0)} + (u/v) \underline f^{(1)}$ and keep only the zeroth order in $(u/v)$ in this section. Upon equating the full collision integral $\underline{St}_{\rm inj}+\underline{St}^{(2)}+\underline{St}^{(3a)}$ to zero we obtain
\begin{align}
&\frac{3}{2} \underline \Gamma_s \underline f^{(0)} + \underline \Gamma_v \hat k_i \underline f_i^{(0)} + \frac{3}{2} \underline \Gamma_t \hat k_i \hat k_j \underline f_{ij}^{(0)} = \nonumber \\ 
 &\frac{1}{2} \underline \Gamma_{tr} \underline f_{ii}^{(0)} + \hat \epsilon_{\zeta_0} \delta (\omega - 2 v p) \lbrace I^{(0)} + \hat k_i I^{(0)}_i + \hat k_i \hat k_j I^{(0)}_{ij} \rbrace. \label{eq:CollIntzerothO}
\end{align}
The solutions for matrix and vector part immediately follow
\begin{subequations}
\begin{eqnarray}
\hat k_i \hat k_j \underline f_{ij}^{(0)} &=& \frac{2}{3}\underline \Gamma_t^{-1} \underline{\hat e}_{\zeta_0} \delta (\omega - 2v p) \hat k_i \hat k_j I_{ij}^{(0)}, \\
\hat k_i \underline f_{i}^{(0)} &=& \underline \Gamma_v^{-1} \underline{\hat e}_{\zeta_0} \delta (\omega - 2v p) \hat k_i I_{i}^{(0)} . \label{eq:VectorPart}
\end{eqnarray}
We insert these solutions back into the Eq.~\eqref{eq:CollIntzerothO} and then solve for the scalar part
\begin{equation}
\underline f^{(0)} \doteq - \frac{2}{3} \underline{\Gamma}_s^{-1} [\frac{1}{3}\underline \Gamma_{tr} \underline \Gamma_t^{-1} - 1]\underline{\hat e}_{\zeta_0} \delta (\omega - 2 v p) I^{(0)}.
\end{equation}
\end{subequations}
The symbol "$\doteq$" means that this equation is valid only if $\underline f^{(0)}$ is multiplied by $(1,-1)$ from the left. This is, because the scalar symmetric part $(f_{\zeta = +} + f_{\zeta = -})$ corresponds to a zero mode of the collision integral:
\begin{equation}
\underline{\Gamma}_s \left (\begin{array}{c}
1 \\ 
1
\end{array} \right ) = 0.
\end{equation}
The inverse of $\underline{\Gamma}_s^{-1}$ means (symbolically in the subspace of the antisymmetric part $(f_{\zeta = +} - f_{\zeta = -})$)
\begin{equation}
\underline{\Gamma}_s^{-1} \doteq \frac{1}{2 \kappa} \left (\begin{array}{cc}
\frac{1}{\Gamma_-} & 0 \\ 
0 & -\frac{1}{\Gamma_+}
\end{array} \right ).
\end{equation}

\subsection{Static solution to $\mathcal O[(u/v)^1]$}

As we shall show below, for the calculation of the current we only need terms containing an odd number of $\hat k$s in the linear corrections to the distribution function. We denote this assumption by a ``$\doteq$'' in the following equation for the first order terms

\begin{widetext}
\begin{eqnarray}
\underline \Gamma_v \hat k_i \underline f_i^{(1)} + \frac{3}{2} \underline \Gamma_t \underline f_{ijk}^{(1)} \hat k_i \hat k_j \hat k_k &\doteq& \frac{1}{10} \underline \xi \underline \Gamma_{tr} \underline \xi \underline f_{ijk}^{(1)}(\hat k_i \delta_{jk} + \hat k_j \delta_{ki} +\hat k_k \delta_{ij}  ) {+ \frac{3}{2} (\hat k \cdot \hat t) \underline \xi [ \underline \Gamma_s \underline f^{(0)} + \underline \Gamma_t \hat k_i \hat k_j \underline f_{ij}^{(0)} - \frac{1}{3} \underline \Gamma_{tr} \underline f_{ii}^{(0)}] }\notag \\
&&{+ \underline \Gamma_{\rm sk} \epsilon_{ijk} \hat k_i \underline f_j^{(0)} \hat t_k} {+ \underline{\hat e}_{\zeta_0} \delta (\omega - 2 v p)\lbrace \hat k_i I_i^{(1)} + \hat k_i \hat k_j \hat k_k I_{ijk}^{(1)}\rbrace} .\label{eq:firstorderEq}
\end{eqnarray}
\end{widetext}
After some algebra, this leads to 
\begin{subequations}
\begin{equation}
\underline f_{ijk}^{(1)} = 2 \underline \Gamma_t^{-1} \underline \xi \underline{\hat e}_{\zeta_0} \delta (\omega - 2 vp) \hat t_k I_{ij}^{(0)}.
\end{equation}
This solution is inserted back into Eq.~\eqref{eq:firstorderEq} and the solution leads to
\begin{eqnarray}
\underline f_i^{(1)} &=&[(\hat t_i I^{(0)}- 2 \hat t_j I_{ij}^{(0)}) \lbrace\underline \Gamma_v^{-1} (1 - \frac{1}{5} \underline \xi \underline \Gamma_{tr} \underline \xi \underline{\Gamma}_t^{-1}) \underline \xi \underline{\hat e}_{\zeta_0} \rbrace \notag \\
&&+ \epsilon_{ijk}  I_j^{(0)} \hat t_k  \underline \Gamma_v^{-1} \underline \Gamma_{\rm sk}\underline \Gamma_v^{-1} \underline{\hat e}_{\zeta_0}]\delta(\omega - 2 vp ) .
\end{eqnarray}
\end{subequations}

\subsection{Static current response}
The current is defined as 
\begin{eqnarray}
\v j &=& - e \sum_{\xi \zeta} \int_{\v p} [v \zeta (\xi \hat p + \frac{u}{v} \hat t)] f_{\xi, \zeta}(\v p) \notag \\
&=& - e v \sum_{\epsilon_e, \epsilon_h} \int_{\v p} (1, -1) [\hat p \underline \xi + \frac{u}{v}\hat t ] \underline f(\v p) .
\end{eqnarray}

Since we found that distribution function is proportional to $\delta (\omega - 2 vp)$ it is useful to switch to spherical coordinates in $\v p$ space (i.e. we transform back from the $\hat k$ representation). From this equation we readlily see that in the terms of first order in $(u/v)$ of $f(\v p)$ we need only terms which contain odd powers of $\hat p$. 

We will write
\begin{eqnarray}
\underline f(\v p) &\doteq &  \delta (\omega - 2 v p ) \Bigg \lbrace \underline F^{(0)} + \hat p_i \underline F^{(0)}_i + \hat  p_i \hat p_j \underline F^{(0)}_{ij} \notag \\
&&+ \frac{u}{v} \left ( \hat p_i \underline F^{(1)}_i + \hat p_i \hat p_j \hat p_k \underline F^{(1)}_{ijk}\right )\Bigg \rbrace.
\end{eqnarray}
Here we introduced
\begin{eqnarray}
\underline F^{(0)} &=&  - \frac{2}{3} \underline{\Gamma}_s^{-1} [\frac{1}{3}\underline \Gamma_{tr} \underline \Gamma_t^{-1} - 1]\underline{\hat e}_{\zeta_0} I^{(0)},\\
\underline F^{(0)}_i &=&  \underline \Gamma_v^{-1} \underline{\hat e}_{\zeta_0}  I_{i}^{(0)}, \\
\underline F^{(0)}_{ij} &=& \frac{2}{3}\underline \Gamma_t^{-1} \underline{\hat e}_{\zeta_0} I_{ij}^{(0)} ,
\end{eqnarray}
as well as 
\begin{eqnarray}
\underline F^{(1)}_i &=& \hat t_i I^{(0)} \underline \Gamma_v^{-1} (1 - \frac{1}{5} \underline \xi \underline \Gamma_{tr} \underline \xi \underline{\Gamma}_t^{-1}) \underline \xi \underline{\hat e}_{\zeta_0} \notag \\
&& + 2 \hat t_j I^{(0)}_{ij}[\frac{2}{3}\underline \Gamma_t^{-1} - \underline \Gamma_v^{-1} (1 - \frac{1}{5} \underline \xi \underline \Gamma_{tr} \underline \xi \underline{\Gamma}_t^{-1})] \underline \xi \underline{\hat e}_{\zeta_0}\notag \\
&&+ \epsilon_{ijk}  I_j^{(0)} \hat t_k  \underline \Gamma_v^{-1} \underline \Gamma_{\rm sk}\underline \Gamma_v^{-1} \underline{\hat e}_{\zeta_0}, \\
\underline F^{(1)}_{ijk} &=& \frac{2}{3}\underline \Gamma_t^{-1} \underline \xi \underline{\hat e}_{\zeta_0}  \hat t_k I_{ij}^{(0)}.
\end{eqnarray}
We used that both $\underline \Gamma_t$ and $\underline \xi$ are diagonal matrices.

In order to present the final result for the current we define $\tilde{\underline \Gamma}_v = \underline \xi {\underline \Gamma}_v \underline \xi$
\begin{align}
\underline M_{\rm sc} &= \tilde{\underline \Gamma}_v^{-1}  -\frac{1}{5} \tilde{\underline \Gamma}_v^{-1} \underline \Gamma_{tr} \underline{\Gamma}_t^{-1} -  \frac{4}{5} \underline \Gamma_t^{-1}- \frac{2}{3}\underline \Gamma_s^{-1} \underline \Gamma_{tr} \underline \Gamma_t^{-1} + 2 \underline \Gamma_s^{-1} , \\
\underline M_{\rm t} &= 2 [ -\tilde{\underline \Gamma}_v^{-1}+ \frac{1}{5} \tilde{\underline \Gamma}_v^{-1}\underline \Gamma_{tr}  \underline{\Gamma}_t^{-1} +  \frac{4}{5} \underline \Gamma_t^{-1} ].
\end{align}
Then we obtain using $\tilde{\underline \Gamma}_{\rm sk} = \underline \xi {\underline \Gamma}_{\rm sk} \underline \xi$
\begin{eqnarray}
j _i &=& - \frac{e^3}{12\pi} \sum_{\epsilon_e \epsilon_h} \frac{1}{2}\Bigg \lbrace (1,-1) \tilde{\underline \Gamma}_v^{-1}  \underline{\hat e}_{\zeta_0} [-i (\boldsymbol{\mathcal E}_+ \times\boldsymbol{\mathcal E}_-)_i] \notag \\
&& +\frac{u}{v} \Bigg [  (1,-1) \tilde{\underline \Gamma}_v^{-1}\tilde{ \underline \Gamma}_{\rm sk}\tilde{\underline \Gamma}_v^{-1} \underline{\hat e}_{\zeta_0}\epsilon_{ijk}  [-i (\boldsymbol{\mathcal E}_+ \times\boldsymbol{\mathcal E}_-)_j] \hat t_k \notag \\
&&+(1,-1)\underline M_{\rm sc} \underline \xi \underline \kappa_3 \underline{\hat e}_{\zeta_0} \frac{\hat t_i \v E(t)^2}{2}  \notag \\
&&- (1,-1)\underline M_{\rm t} \underline \xi \underline \kappa_3\underline{\hat e}_{\zeta_0} \v E_i (t)\frac{(\v E(t) \cdot \hat t)}{2} \Bigg ]\Bigg \rbrace .
\end{eqnarray}
Here $\underline \kappa_3 = \text{diag}(1,-1)_{\zeta}$. Bear in mind that the second harmonic term should not be inferred from this term. The prefactor ${{e^3}/{12 \pi} }$ corresponds to the quantization unit presented in Ref.~\onlinecite{deJuanMoore2017}. We can readily read off the relaxation times introduced in Eq.~\eqref{eq:SteadyStatePGE}
\begin{subequations}
\begin{eqnarray}
&&\tau_{\rm CPGE} = \frac{1}{2}\sum_{\epsilon = \epsilon_e \epsilon_h} (1,-1) \tilde{\underline \Gamma}_v^{-1} \underline{\hat e}_{\zeta_0}, \\
&&\tau_{\rm skew} = \frac{1}{2}\sum_{\epsilon = \epsilon_e, \epsilon_h} (1,-1) \tilde{\underline \Gamma}_v^{-1}\tilde{\underline \Gamma}_{\rm sk}\tilde{\underline \Gamma}_v^{-1} \underline{\hat e}_{\zeta_0} ,\\
&&\tau_{\hat t} = -\frac{1}{2}\sum_{\epsilon = \epsilon_e, \epsilon_h} (1,-1) {\underline M}_{\rm sc} \underline \xi \underline\kappa_3 \underline{\hat e}_{\zeta_0}, \\
&&\tau_{\perp} =-\frac{1}{2}\sum_{\epsilon = \epsilon_e, \epsilon_h} (1,-1) {\underline M}_{\rm t} \underline \xi\underline \kappa_3 \underline{\hat e}_{\zeta_0}.
\end{eqnarray}
\end{subequations}

The evaulation of these equations in the case $\zeta_0 = +1$ leads to
\begin{subequations}
\label{eq:FullRelaxTimes}
\begin{widetext}
\begin{eqnarray}
&&\tau_{\rm CPGE}= \frac{1}{2}\sum_{\epsilon = \epsilon_e, \epsilon_h} \frac{\Gamma_- + \Gamma_+ \kappa}{\Gamma_- \Gamma_+ (1 + 2 \kappa^2) + \frac{3 \kappa}{2}(\Gamma_-^2+\Gamma_+^2)} ,\\
&&\tau_{\rm skew} = \frac{1}{2}\sum_{\epsilon = \epsilon_e, \epsilon_h} (1-\kappa)\frac{ \Gamma_{\text{sk},+}\Gamma_-\left [\Gamma_-+ \Gamma_+ \frac{(3 - \kappa)\kappa}{2}\right ] + \Gamma_{\text{sk},-}\Gamma_+ \kappa \left (\Gamma_- - \Gamma_+ \frac{1- 3 \kappa}{2}\right )}{[\Gamma_- \Gamma_+ (1 + 2 \kappa^2) + \frac{3 \kappa}{2}(\Gamma_-^2+\Gamma_+^2)]^2} , \\
&&\tau_{\hat t} = -\frac{1}{2}\sum_{\epsilon = \epsilon_e, \epsilon_h} \left [ \frac{2}{3 \Gamma_- \kappa }+\frac{4}{15 (\Gamma_++\Gamma_- \kappa )}+\frac{2 (\Gamma_-+\Gamma_+ \kappa )}{5 [\Gamma_- \Gamma_+ (1 + 2 \kappa^2) + \frac{3 \kappa}{2}(\Gamma_-^2+\Gamma_+^2)]}\right  ] \text{sgn}\left (\epsilon-{\gamma}\right) ,\\
&&\tau_{\perp } =-\frac{1}{2}\sum_{\epsilon = \epsilon_e, \epsilon_h}\kappa\left [\frac{2   \left(\Gamma_-^2+\Gamma_+^2+2 \Gamma_- \Gamma_+ \kappa \right)}{5 (\Gamma_++\Gamma_- \kappa ) [\Gamma_- \Gamma_+ (1 + 2 \kappa^2) + \frac{3 \kappa}{2}(\Gamma_-^2+\Gamma_+^2)]}\right ]\text{sgn}\left (\epsilon-{\gamma}\right).
\end{eqnarray}
\end{widetext}
\end{subequations}
We note that the two relaxation times $\tau_{\hat t}$ and $\tau_{\perp}$ contain additional factors of $\sgn{\epsilon - \gamma/2}$ and therefore have opposite signs for photoelectrons and photoholes. 

\subsection{Dynamic solution to $\mathcal O[(u/v)^0]$}

In order to obtain the time dependence of the photocurrent after a sudden illumination, we add the solution of the homogeneous Boltzmann equation to our steady state solution, Eq.~\eqref{eq:VectorPart}. Under the assumption of thermal equilibrium at $t = 0$ we obtain.
\begin{equation}
\hat k_i \underline{f}_i= (1- e^{- \underline{ \Gamma}_v t}) \underline \Gamma_v^{-1} \underline{\hat e}_{\zeta_0} \delta (\omega - 2v p) \hat k_i I_{i}^{(0)} .
\end{equation}
Inserting this into the definition of the current leads to $\tau_{\rm CPGE} \rightarrow T(t)$ with
\begin{equation}
T(t) = \frac{1}{2} \sum_{\epsilon = \epsilon_e, \epsilon_h}(1,-1) (1- e^{- \underline{ \tilde \Gamma}_v t})  \tilde{\underline \Gamma}_v^{-1} \underline{\hat e}_{\zeta_0} .
\end{equation}
We assume that $\zeta_0 = +1$ and find, after some algebra, an expression by means of $\overline \Gamma = \frac{\Gamma_+ + \Gamma_-}{2}$ and $\Delta \Gamma = \Gamma_+ - \Gamma_-$ to be
\begin{widetext}
\begin{eqnarray}
T(t) = \frac{1}{2} \sum_{\epsilon = \epsilon_e, \epsilon_h}\sum_{\pm} \Bigg \lbrace \frac{1-\exp\left[-(t/2)\left[2 \overline \Gamma + 3\kappa \overline \Gamma \pm \sqrt{(\Delta \Gamma)^2(1- \kappa)( 1-2 \kappa) + \overline \Gamma^2 \kappa^2}\right]\right]}{\left[2\overline \Gamma + 3\kappa \overline \Gamma \pm \sqrt{(\Delta \Gamma)^2(1- \kappa)( 1-2 \kappa) + \overline \Gamma^2 \kappa^2}\right]} \nonumber \\ \times\left [1 \pm \frac{\Delta \Gamma (1- \kappa) + \overline \Gamma \kappa}{\sqrt{(\Delta \Gamma)^2(1- \kappa)( 1-2 \kappa) + \overline \Gamma^2 \kappa^2}}\right ]\Bigg \rbrace. \label{eq:TdepGeneral} 
\end{eqnarray}
\end{widetext}
It is worth to notice, that $\Big [1 - \frac{\Delta \Gamma (1- \kappa) + \overline \Gamma \kappa}{\sqrt{(\Delta \Gamma)^2(1- \kappa)( 1-2 \kappa) + \overline \Gamma^2 \kappa^2}}\Big ]$ changes sign at $\Delta \Gamma = 0$. This is the origin of the non-monotonic behavior reported in Fig.~\ref{fig:TimeEvolution}.

\section{Intraband response}
\label{app:intraband}

Here we present details on the contribution from the Fermi surface. We concentrate on the limit $u/v = 0$. The side jump contribution modifies the collision integral as follows
\begin{widetext}
\begin{eqnarray}
St_\zeta[f_+, f_-] &=& -\sum \hspace{-.4cm} \int_{l'} 2 \pi \vert \braket{u_{\xi, \v p} \vert u_{\xi', \v p'}} \vert ^2 n_{\rm imp} \vert \mathcal{V}_{(\zeta - \zeta') \bm b} \vert ^2 \delta \left (\epsilon_{l'} - \epsilon_l - e \v E \delta \v r_{l'l}\right ) \left [ \tilde f_{l}(\epsilon_{l}, \hat p, t)-\tilde f_{l'}(\epsilon_{l'}, \hat p', t)\right ] \notag \\
&=& - 2 \pi n_{\rm imp} \vert \mathcal{V}_0 \vert ^2 \Big \langle \vert \braket{u_{\xi, \v p} \vert u_{\xi', \v p'}} \vert ^2 \nu_\zeta (\epsilon_l + e \v E \delta \v r_{l'l})  \left [\tilde f_\zeta (\epsilon_l, \hat p, t) - \tilde f_\zeta (\epsilon_l + e \v E \delta \v r_{l'l} , \hat p', t\right ]\Big \rangle_{\hat p'} \notag \\
&&- 2 \pi n_{\rm imp} \vert \mathcal{V}_{2 \bm b} \vert ^2 \Big \langle \vert \braket{u_{\xi, \v p} \vert u_{\xi', \v p'}} \vert ^2 \nu_{-\zeta} (\epsilon_l + e \v E \delta \v r_{l'l}) \left [\tilde f_\zeta (\epsilon_l, \hat p, t) - \tilde f_{-\zeta} (\epsilon_l + e \v E \delta \v r_{l'l} , \hat p', t\right ]\Big \rangle_{\hat p'}.
\end{eqnarray}
\end{widetext}
Here, $\tilde f_{\zeta} (\epsilon, \hat p) = f_\zeta (\v p)$ and the sidejump is 
\begin{eqnarray}
\delta \v r_{l'l} &=& \boldsymbol{\mathcal A}_{l'} -\boldsymbol{\mathcal A}_{l} - (\partial_{\v p} + \partial_{\v p'}) \arg\left (\braket{l'\vert l}\right ) \notag \\
&=& - \frac{\hat p \times \hat p'}{4 \vert \braket{u_{\xi,\v p} \vert u_{\xi',\v p'}} \vert^2}\left \lbrace \frac{\xi}{p'} + \frac{\xi'}{p}\right \rbrace.
\end{eqnarray}
In this appendix, we investigate how the driving term proportional to $\dot{\v p}$ excites particle-hole pairs at the Fermi surface. We concentrate on the circular PGE. The according contribution to the distribution function is
\begin{align}
\underline f &= f_{\rm FD} \left (\begin{array}{c}
1/2 \\ 
1/2
\end{array} \right ) +\hat p\cdot \boldsymbol{\mathcal E} e^{i \omega t} ev f'_{FD}(\epsilon_l) [i \underline{\Omega}(\mu)]^{-1} \left (\begin{array}{c}
\xi_+ \\ 
-\xi_-
\end{array} \right ) \notag \\
&- \hat p \cdot (\boldsymbol{\mathcal E}\times\boldsymbol{\mathcal E}^*)  v e^2 \underline{M}^{-1}\underline N \lbrace \underline{\nu}\left [i \underline{\Omega}\right ]^{-1}f'_{FD}\rbrace' \left (\begin{array}{c}
\xi_+ \\ 
-\xi_-
\end{array} \right ) + c.c..
\end{align}
Here, $\underline M = \underline \Gamma_v/(2\pi n_{\rm imp})$, $i\underline \Omega= i \omega + \underline \Gamma_v$ and 
\begin{equation}
\underline N = {\frac{-v}{6}\left (\begin{array}{cc}
\frac{\vert \mathcal{V}_0\vert^2 }{\gamma-\epsilon_l} & \frac{\xi_+ \xi_-\gamma \vert \mathcal{V}_{2 \bm b} \vert ^2}{\gamma^2-\epsilon_l^2} \\ 
\frac{\xi_+ \xi_-\gamma \vert \mathcal{V}_{2 \bm b} \vert ^2}{\gamma^2-\epsilon_l^2} & \frac{\vert \mathcal{V}_0\vert^2 }{\gamma+\epsilon_l}
\end{array} \right )}
\end{equation}
accounts for the side jump contribution. When evaluated by means of the definition of the current, which however now contains additional contributions:
\begin{equation}
\v j = - e \sum_{\xi, \zeta}\int_{\v p} v [\zeta \xi \hat p - e \boldsymbol{\Omega}_\xi \times \v E + \delta \dot{\v r}_{\xi \zeta}] f_\zeta(\v p).
\end{equation}
In the present case, the side jump accumulation is zero
\begin{equation}
\delta \dot{\v r}_{\zeta \xi} = \sum\hspace{-.4cm} \int_{l'} w_{l'l} \delta \v r_{l'l} = 0.
\end{equation}
We obtain in linear response
\begin{equation}
\v j_{\rm lin} = e^2 \frac{v^2}{3} \boldsymbol{\mathcal E}e^{i \omega t}\left (\xi_+, -\xi_-\right ) \underline \nu [i \underline \Omega]^{-1}\left (\begin{array}{c}
\xi_+ \\ 
-\xi_-
\end{array} \right )  +c.c..
\end{equation}
The second order response contains two contributions. First, the Berry curvature term leads to
\begin{equation}
\v j_{\rm intr} = \frac{e^3}{4 \pi^2} \frac{\boldsymbol{\mathcal E} \times\boldsymbol{\mathcal E}^*}{3} \left (\xi_+, \xi_-\right ) [i \underline \Omega]^{-1}\left (\begin{array}{c}
\xi_+ \\ 
-\xi_-
\end{array} \right ) + c.c..
\end{equation}
Clearly, in the clean limit contributions from opposite cones cancel up.
Furthermore, there is a contribution from the usual group velocity combined with the side jump term
\begin{widetext}
\begin{align}
&\v j_{\rm sj} = v^2 e^3 \frac{\boldsymbol{\mathcal E} \times\boldsymbol{\mathcal E}^*}{3} \left (\xi_+,-\xi_- \right )[\underline \nu \underline{M}^{-1}\underline N]' \underline{\nu}[i \underline{\Omega}]^{-1} \left (\begin{array}{c}
\xi_+ \\ 
-\xi_-
\end{array} \right ) +c.c. \notag \\
&= \frac{-v^3}{2 \gamma} e^3 \frac{\boldsymbol{\mathcal E} \times\boldsymbol{\mathcal E}^*}{3} \left (\xi_+,-\xi_- \right )[(\mathbf 1 + \kappa \underline R)^{-1} \underline S]' \underline{\nu} [i \underline{\Omega}]^{-1} \left (\begin{array}{c}
\xi_+ \\ 
-\xi_-
\end{array} \right ) + c.c. .
\end{align}
\end{widetext}
In the second line, we introduced the dimensionless matrices 
\begin{align}
\underline R(\epsilon_F) &= \frac{1}{2} \left (\begin{array}{cc}
2 \left (\frac{\epsilon_F + \gamma}{\epsilon_F - \gamma}\right)^2 & - \xi_+ \xi_- \\ 
- \xi_+ \xi_- & 2 \left (\frac{\epsilon_F - \gamma}{\epsilon_F + \gamma}\right)^2
\end{array} \right ) ,\\
\underline S(\epsilon_F) &= \left (\begin{array}{cc}
\frac{\gamma}{\gamma - \epsilon_F} & \xi_+ \xi_- \kappa \frac{\gamma^2}{\gamma^2 - \epsilon_F^2}  \\
 \xi_+ \xi_- \kappa \frac{\gamma^2}{\gamma^2 - \epsilon_F^2} &\frac{\gamma}{\gamma + \epsilon_F}
\end{array} \right ).
\end{align}
Keep in mind that the prime $'$ denotes derivative with respect to energy. In the limit $\kappa \rightarrow 0$ (no intervalley scattering) we have $\frac{v^3}{\gamma}[(\mathbf 1 + \kappa \underline R)^{-1} \underline S]' \underline \nu \rightarrow \text{diag}(1,-1)/[2 \pi^2]$ so that 
\begin{equation}
\v j_{\rm sj} \stackrel{\kappa \rightarrow 0}{\longrightarrow} - \v j_{\rm intr}.
\end{equation}
Remarkably, the contribution from the side jump exactly cancels up the intrinsic contribution. Generally, we can state the the Fermi surface contributions are $\Gamma_\zeta/\omega \ll 1$ times smaller than the contributions from photocarriers.

\section{Microscopic Model}
\label{app:MicroModel}

As the simplest model of the Weyl semimetal (WSM), we review the two-band multilayer Chern insulator Hamiltonian [\onlinecite{ShapourianHughes2016,deJuanMoore2017}]. The disorder-free single-particle Hamiltonian on the cubic lattice reads
\begin{subequations} \label{l-ham}
\begin{align}
H_0 = &\, \frac{t}{2} \sum_{\bfr}\sum_{s=1,2}\left[ C_{\bfr+\bfa_s}^\dagger \left( i \sigma_s - \sigma_3  \right) C_\bfr + \mathrm{H.c.}\right] \notag \\
&\, + \sum_{\bfr}{ C_\bfr^\dagger \left( M \, \sigma_3 -\mu \right) C_\bfr} \label{l-ham-1}  \\
      &\, +\frac{1}{2}\sum_{\bfr}{ \left[ C_{\bfr+\bfa_3}^\dagger \left( i \gamma - t \, \sigma_3 \right) C_\bfr + \mathrm{H.c.} \right]} ,  \label{l-ham-2}
\end{align}
\end{subequations}
where $\bfr = (x \, \, y \, \, z)$ is the coordinate of a site on the cubic lattice, $(\bfa_1 \, \, \bfa_2 \,\, \bfa_3) = (\hat{\mathbf{x}} \, \, \hat{\mathbf{y}} \, \, \hat{\mathbf{z}})$ the set of unit vectors, $C_{\bfr}^\dagger \equiv [c_\uparrow^\dagger(\bfr) \,\, c_\downarrow^\dagger(\bfr)]$ ($C_\bfr$) the creation (annihilation) operator of an electron spinor, and ${\bsig} = (\sigma_1, \, \sigma_2, \, \sigma_3 )$ the Pauli matrices in the spin space.  The Hamiltonian (\ref{l-ham}) describes a stack of Chern insulator layers [Eq.~(\ref{l-ham-1})] coupled through tunneling [Eq.~(\ref{l-ham-2})] and all the model parameters $t,\gamma,M$, and $\mu$ are real. 

In reciprocal space the Hamiltonian takes the form 
\be
H_0 = \sum_{\bk}{\psi^\dagger(\bk) \, \mathcal{H}_{0}(\bk) \, \psi(\bk)}, 
\ee
with $\psi(\bk)$ the two-spinor
\be \label{2sp-0}
\psi(\bk) = \begin{bmatrix} c_\uparrow(\bk) \\  c_\downarrow(\bk) \end{bmatrix}, 
\ee
and the Hamiltonian matrix 
\begin{subequations} \label{Ham0}
\begin{align}
& \mathcal{H}_{0}(\bk) =  \vep_{\bk} -\mu  + \bd_{\bk} \cdot \hat{\bsig}, \\
& \vep_{\bk} = \gamma \sin{k_z}, \\
& \bd_{\bk} = \big( t\sin{k_x}, \,\, t\sin{k_y}, \,\, -M+ t \sum_{j = x,y,z}{\cos{k_j}} \big).
\end{align}
\end{subequations} 
The energy dispersion is 
\begin{widetext}
\be
\begin{split}
E_{\pm}(\bk) = &\, \vep_{\bk} -\mu  \\ 
&\, \pm \sqrt{M^2 - 2 M t \left( \sum_{j = x,y,z}{\cos{k_i}} \right) + 2 \, t^2 \left(\cos{k_x}\cos{k_y} + \cos{k_x}\cos{k_z} + \cos{k_y}\cos{k_z} + \frac{1}{2}\cos^2{k_z} + 1 \right)},
\end{split}
\ee
\end{widetext}
For $|M/t-2|<1$, a pair of Weyl valleys form about the momenta and energies
\begin{subequations}
\begin{align}
\bK_{\zeta} & \, = \left(0, \, 0, \, \zeta K \right), \quad K =\cos^{-1}(M/t-2), \\
E_{\zeta}  & \, = \zeta E - \mu, \quad E = \gamma \sin{K},
\end{align}
\end{subequations} 
where $\zeta = \pm 1$ indicates the valleys. For $|M/t-2|=1$ there is only one band-touching point and for $|M/t-2|>1$ band gap opens, for which the physics is less interesting. 

In the Weyl semimetal regime $|M/t-2|<1$, we expand the Hamiltonian (\ref{Ham0}) about the Weyl point $\bK_{\zeta}$ to obtain a low-energy continuous model. Substituting $\bk = \bK_{\zeta} + \bp$ into Eq.~(\ref{Ham0}), for small momentum deviation $|\bp| \ll K$ and up to linear order in $p_{x,y,z}$ we have  
\begin{align} \label{low-e-ham-0}
\mathcal{H}_{0}(\bK_{\zeta} + \bp) & \simeq E_{\zeta} + \cos{K} \, p_z \notag \\
&+ t \, \left( p_x \sigma_1 +p_y \sigma_2 + \zeta \sin{K} \, p_z \sigma_3 \right),
\end{align}
and the corresponding low-energy two-spinor (\ref{2sp-0}) takes the valley index $\zeta$ (chirality)
\be
\psi_\zeta(\bp) \equiv  \begin{bmatrix} \psi_{\uparrow,\zeta}(\bp) \\  \psi_{\downarrow,\zeta}(\bp) \end{bmatrix} \simeq \begin{bmatrix} c_{\uparrow}(\bK_{\zeta} + \bp) \\  c_{\downarrow}(\bK_{\zeta}+\bp) \end{bmatrix}.
\ee
Introducing the four-spinor
\be 
\Psi(\bp) \equiv  \begin{pmatrix} 1 & 0 \\ 0 &   \sigma_z \end{pmatrix} \begin{bmatrix} \psi_\mathrm{+1}(\bp) \\ \psi_{-1}(\bp) \end{bmatrix},
\ee 
 and the Pauli matrices ${\boldsymbol{\k}} = (\kappa_1, \, \kappa_2, \, \kappa_3 )$ for valley space, we obtain the Weyl Hamiltonian 
\begin{align} \label{lowE-h0-0}
H_{\rm kin}(\bp) &=  E \, \kappa_3   + \cos{K} \, p_z  \notag \\
&+ t \, \kappa_3 \left(  p_x \sigma_1 +p_y \sigma_2 + \sin{K} \, p_z \sigma_3 \right).
\end{align}
Especially, at $M/t=2$, where $K = \pi/2$ and $E = \gamma$, the Weyl cones become isotropic and the Hamiltonian (\ref{lowE-h0-0}) takes the form of Eq.~\eqref{eq:Hkin} at $u/v= 0$.



\begin{thebibliography}{27}%
\makeatletter
\providecommand \@ifxundefined [1]{%
 \@ifx{#1\undefined}
}%
\providecommand \@ifnum [1]{%
 \ifnum #1\expandafter \@firstoftwo
 \else \expandafter \@secondoftwo
 \fi
}%
\providecommand \@ifx [1]{%
 \ifx #1\expandafter \@firstoftwo
 \else \expandafter \@secondoftwo
 \fi
}%
\providecommand \natexlab [1]{#1}%
\providecommand \enquote  [1]{``#1''}%
\providecommand \bibnamefont  [1]{#1}%
\providecommand \bibfnamefont [1]{#1}%
\providecommand \citenamefont [1]{#1}%
\providecommand \href@noop [0]{\@secondoftwo}%
\providecommand \href [0]{\begingroup \@sanitize@url \@href}%
\providecommand \@href[1]{\@@startlink{#1}\@@href}%
\providecommand \@@href[1]{\endgroup#1\@@endlink}%
\providecommand \@sanitize@url [0]{\catcode `\\12\catcode `\$12\catcode
  `\&12\catcode `\#12\catcode `\^12\catcode `\_12\catcode `\%12\relax}%
\providecommand \@@startlink[1]{}%
\providecommand \@@endlink[0]{}%
\providecommand \url  [0]{\begingroup\@sanitize@url \@url }%
\providecommand \@url [1]{\endgroup\@href {#1}{\urlprefix }}%
\providecommand \urlprefix  [0]{URL }%
\providecommand \Eprint [0]{\href }%
\providecommand \doibase [0]{http://dx.doi.org/}%
\providecommand \selectlanguage [0]{\@gobble}%
\providecommand \bibinfo  [0]{\@secondoftwo}%
\providecommand \bibfield  [0]{\@secondoftwo}%
\providecommand \translation [1]{[#1]}%
\providecommand \BibitemOpen [0]{}%
\providecommand \bibitemStop [0]{}%
\providecommand \bibitemNoStop [0]{.\EOS\space}%
\providecommand \EOS [0]{\spacefactor3000\relax}%
\providecommand \BibitemShut  [1]{\csname bibitem#1\endcsname}%
\let\auto@bib@innerbib\@empty

\bibitem [{\citenamefont {Burkov}(2016)}]{Burkov2016}%
  \BibitemOpen
  \bibfield  {author} {\bibinfo {author} {\bibfnamefont {A.~A.}\ \bibnamefont
  {Burkov}},\ }\href@noop {} {\bibfield  {journal} {\bibinfo  {journal} {Nature
  Materials}\ }\textbf {\bibinfo {volume} {15}},\ \bibinfo {pages} {1145}
  (\bibinfo {year} {2016})}\BibitemShut {NoStop}%
\bibitem [{\citenamefont {Weyl}(1929)}]{Weyl1929}%
  \BibitemOpen
  \bibfield  {author} {\bibinfo {author} {\bibfnamefont {H.}~\bibnamefont
  {Weyl}},\ }\href@noop {} {\bibfield  {journal} {\bibinfo  {journal}
  {Zeitschrift f{\"u}r Physik}\ }\textbf {\bibinfo {volume} {56}},\ \bibinfo
  {pages} {330} (\bibinfo {year} {1929})}\BibitemShut {NoStop}%
\bibitem{NeutrinoOscillations}%
Y.~Fukuda \textit{et al.} (Super-Kamiokande collaboration), Phys. Rev. Lett. \textbf{81}, 1562 (1998); Q.~R.~Ahmad \textit{et al.} (SNO collaboration), Phys. Rev. Lett. \textbf{87}, 071301 (2001)
\bibitem [{\citenamefont {Bertlmann}(2000)}]{Bertlmann2000}%
  \BibitemOpen
  \bibfield  {author} {\bibinfo {author} {\bibfnamefont {R.}~\bibnamefont
  {Bertlmann}},\ }\href@noop {} {\emph {\bibinfo {title} {Anomalies in Quantum
  Field Theory}}},\ International Series of Monographs on Physics\ (\bibinfo
  {publisher} {Clarendon Press},\ \bibinfo {year} {2000})\BibitemShut {NoStop}%
\bibitem [{\citenamefont {Son}\ and\ \citenamefont
  {Spivak}(2013)}]{SonSpivak2013}%
  \BibitemOpen
  \bibfield  {author} {\bibinfo {author} {\bibfnamefont {D.}~\bibnamefont
  {Son}}\ and\ \bibinfo {author} {\bibfnamefont {B.}~\bibnamefont {Spivak}},\
  }\href@noop {} {\bibfield  {journal} {\bibinfo  {journal} {Physical Review
  B}\ }\textbf {\bibinfo {volume} {88}},\ \bibinfo {pages} {104412} (\bibinfo
  {year} {2013})}\BibitemShut {NoStop}%
\bibitem [{\citenamefont {Xiong}\ \emph {et~al.}(2015)\citenamefont {Xiong},
  \citenamefont {Kushwaha}, \citenamefont {Liang}, \citenamefont {Krizan},
  \citenamefont {Hirschberger}, \citenamefont {Wang}, \citenamefont {Cava},\
  and\ \citenamefont {Ong}}]{XiongOng2015}%
  \BibitemOpen
  \bibfield  {author} {\bibinfo {author} {\bibfnamefont {J.}~\bibnamefont
  {Xiong}}, \bibinfo {author} {\bibfnamefont {S.~K.}\ \bibnamefont {Kushwaha}},
  \bibinfo {author} {\bibfnamefont {T.}~\bibnamefont {Liang}}, \bibinfo
  {author} {\bibfnamefont {J.~W.}\ \bibnamefont {Krizan}}, \bibinfo {author}
  {\bibfnamefont {M.}~\bibnamefont {Hirschberger}}, \bibinfo {author}
  {\bibfnamefont {W.}~\bibnamefont {Wang}}, \bibinfo {author} {\bibfnamefont
  {R.}~\bibnamefont {Cava}}, \ and\ \bibinfo {author} {\bibfnamefont
  {N.}~\bibnamefont {Ong}},\ }\href@noop {} {\bibfield  {journal} {\bibinfo
  {journal} {Science}\ }\textbf {\bibinfo {volume} {350}},\ \bibinfo {pages}
  {413} (\bibinfo {year} {2015})}\BibitemShut {NoStop}%
\bibitem{WeylRealizations}%
Huang S. M. \textit{et al.}, Nat. Commun. \textbf{6}, 7373 (2015); 
Xu S.-Y. \textit{et al.}, Science \textbf{349}, 613–617 (2015); 
Lv B. Q. \textit{et al.},  Phys. Rev. X \textbf{5}, 031013 (2015); 
Ch.~Shekhar \textit{et al.}, Nat. Phys. \textbf{11}, 645–649 (2015);
B.~Q.~Lv \textit{et al.}, Nat. Phys. \textbf{11}, 724–727 (2015); 
L.-X.~Yang \textit{et al.}, Nat. Phys. \textbf{11}, 728–732 (2015); 
Xu S.-Y. \textit{et al.}, Nat. Phys. \textbf{11}, 748–754 (2015); 
Xu S.-Y. \textit{et al.}, Science Advances \textbf{1}, e1501092 (2015);
N.~ Xu \textit{et al.}, Nat. Commun. \textbf{7}, 11006 (2016);
Liu Z. K. \textit{et al.}, Nat. Mater. \textbf{15}, 27–31 (2016).
\bibitem [{Cha()}]{ChanLee2016}%
C.-K.~Chan, N.~Lindner, G.~Refael and P.~Lee, {Phys. Rev. B} \textbf{95},
{041104} ({2017})
\bibitem [{\citenamefont {Wu}\ \emph {et~al.}(2016)\citenamefont {Wu},
  \citenamefont {Patankar}, \citenamefont {Morimoto}, \citenamefont {Nair},
  \citenamefont {Thewalt}, \citenamefont {Little}, \citenamefont {Analytis},
  \citenamefont {Moore},\ and\ \citenamefont {Orenstein}}]{WuOrenstein2016}%
  \BibitemOpen
  \bibfield  {author} {\bibinfo {author} {\bibfnamefont {L.}~\bibnamefont
  {Wu}}, \bibinfo {author} {\bibfnamefont {S.}~\bibnamefont {Patankar}},
  \bibinfo {author} {\bibfnamefont {T.}~\bibnamefont {Morimoto}}, \bibinfo
  {author} {\bibfnamefont {N.~L.}\ \bibnamefont {Nair}}, \bibinfo {author}
  {\bibfnamefont {E.}~\bibnamefont {Thewalt}}, \bibinfo {author} {\bibfnamefont
  {A.}~\bibnamefont {Little}}, \bibinfo {author} {\bibfnamefont {J.~G.}\
  \bibnamefont {Analytis}}, \bibinfo {author} {\bibfnamefont {J.~E.}\
  \bibnamefont {Moore}}, \ and\ \bibinfo {author} {\bibfnamefont
  {J.}~\bibnamefont {Orenstein}},\ }\href@noop {} {} (\bibinfo {year} {2016}),\
  \bibinfo {note} {(advance online publication)}\BibitemShut {NoStop}%
\bibitem [{\citenamefont {Morimoto}\ \emph {et~al.}(2016)\citenamefont
  {Morimoto}, \citenamefont {Zhong}, \citenamefont {Orenstein},\ and\
  \citenamefont {Moore}}]{MorimotoMoore2016}%
  \BibitemOpen
  \bibfield  {author} {\bibinfo {author} {\bibfnamefont {T.}~\bibnamefont
  {Morimoto}}, \bibinfo {author} {\bibfnamefont {S.}~\bibnamefont {Zhong}},
  \bibinfo {author} {\bibfnamefont {J.}~\bibnamefont {Orenstein}}, \ and\
  \bibinfo {author} {\bibfnamefont {J.~E.}\ \bibnamefont {Moore}},\ }\href@noop
  {} {\bibfield  {journal} {\bibinfo  {journal} {Phys. Rev. B}\ }\textbf
  {\bibinfo {volume} {94}},\ \bibinfo {pages} {245121} (\bibinfo {year}
  {2016})}\BibitemShut {NoStop}%
  \bibitem{Jarillo-Herrero-Gedik2017}
Qiong Ma, Su-Yang Xu, Ching-Kit Chan, Cheng-Long Zhang, Guoqing Chang, Yuxuan Lin, Weiwei Xie, Tom\'as Palacios, Hsin Lin, Shuang Jia, Patrick A. Lee, Pablo Jarillo-Herrero, Nuh Gedik, preprint arXiv:1705.00590 [to appear in Nature Physics].   
\bibitem [{\citenamefont {de~Juan}\ \emph {et~al.}(2016)\citenamefont
  {de~Juan}, \citenamefont {Grushin}, \citenamefont {Morimoto},\ and\
  \citenamefont {Moore}}]{deJuanMoore2017}%
  \BibitemOpen
  \bibfield  {author} {\bibinfo {author} {\bibfnamefont {F.}~\bibnamefont
  {de~Juan}}, \bibinfo {author} {\bibfnamefont {A.~G.}\ \bibnamefont
  {Grushin}}, \bibinfo {author} {\bibfnamefont {T.}~\bibnamefont {Morimoto}}, \
  and\ \bibinfo {author} {\bibfnamefont {J.~E.}\ \bibnamefont {Moore}},\
  }\href@noop {} {\bibfield  {journal} {\bibinfo  {journal} {arXiv preprint
  arXiv:1611.05887}\ } (\bibinfo {year} {2016})}\BibitemShut {NoStop}%
\bibitem [{\citenamefont {Belinicher}\ and\ \citenamefont
  {Sturman}(1980)}]{BelinicherSturman1980}%
  \BibitemOpen
  \bibfield  {author} {\bibinfo {author} {\bibfnamefont {V.~I.}\ \bibnamefont
  {Belinicher}}\ and\ \bibinfo {author} {\bibfnamefont {B.~I.}\ \bibnamefont
  {Sturman}},\ }\href@noop {} {\bibfield  {journal} {\bibinfo  {journal}
  {Uspekhi Fizicheskih Nauk}\ }\textbf {\bibinfo {volume} {130}},\ \bibinfo
  {pages} {415} (\bibinfo {year} {1980})}\BibitemShut {NoStop}%
\bibitem [{\citenamefont {Sturman}\ and\ \citenamefont
  {Fridkin}(1992)}]{SturmanFridkin1992}%
  \BibitemOpen
  \bibfield  {author} {\bibinfo {author} {\bibfnamefont {B.~I.}\ \bibnamefont
  {Sturman}}\ and\ \bibinfo {author} {\bibfnamefont {V.~M.}\ \bibnamefont
  {Fridkin}},\ }\href@noop {} {\emph {\bibinfo {title} {Photovoltaic and
  Photorefractive Effects in Noncentrosymmetric Materials}}}\ (\bibinfo
  {publisher} {Gordon and Breach},\ \bibinfo {year} {1992})\BibitemShut
  {NoStop}%
\bibitem [{\citenamefont {Ivchenko}(2005)}]{Ivchenko2005}%
  \BibitemOpen
  \bibfield  {author} {\bibinfo {author} {\bibfnamefont {E.}~\bibnamefont
  {Ivchenko}},\ }\href@noop {} {\emph {\bibinfo {title} {Optical Spectroscopy
  of Semiconductor Nanostructures}}}\ (\bibinfo  {publisher} {Alpha Science},\
  \bibinfo {year} {2005})\BibitemShut {NoStop}%
\bibitem [{\citenamefont {Sipe}\ and\ \citenamefont
  {Shkrebtii}(2000)}]{SipeShkrebtii2000}%
  \BibitemOpen
  \bibfield  {author} {\bibinfo {author} {\bibfnamefont {J.~E.}\ \bibnamefont
  {Sipe}}\ and\ \bibinfo {author} {\bibfnamefont {A.~I.}\ \bibnamefont
  {Shkrebtii}},\ }\href@noop {} {\bibfield  {journal} {\bibinfo  {journal}
  {Phys.~Rev.~B}\ }\textbf {\bibinfo {volume} {61}},\ \bibinfo {pages}
  {5337} (\bibinfo {year} {2000})}\BibitemShut {NoStop}%
\bibitem [{\citenamefont {Morimoto}\ and\ \citenamefont
  {Nagaosa}(2016)}]{MorimotoNagaosa2016}%
  \BibitemOpen
  \bibfield  {author} {\bibinfo {author} {\bibfnamefont {T.}~\bibnamefont
  {Morimoto}}\ and\ \bibinfo {author} {\bibfnamefont {N.}~\bibnamefont
  {Nagaosa}},\ }\href@noop {} {\bibfield  {journal} {\bibinfo  {journal}
  {Science Advances}\ }\textbf {\bibinfo {volume} {2}} (\bibinfo {year}
  {2016})}\BibitemShut {NoStop}%
\bibitem [{Koe()}]{Koenig2017}%
  \BibitemOpen
  \href@noop {} {}\bibinfo {note} {E.J.K\"{o}nig \textit{et al., in
  preparation.}}\BibitemShut {Stop}%
\bibitem [{\citenamefont {von Baltz}\ and\ \citenamefont
  {Kraut}(1981)}]{vonBaltzKraut1981}%
  \BibitemOpen
  \bibfield  {author} {\bibinfo {author} {\bibfnamefont {R.}~\bibnamefont {von
  Baltz}}\ and\ \bibinfo {author} {\bibfnamefont {W.}~\bibnamefont {Kraut}},\
  }\href@noop {} {\bibfield  {journal} {\bibinfo  {journal} {Phys. Rev. B}\
  }\textbf {\bibinfo {volume} {23}},\ \bibinfo {pages} {5590} (\bibinfo {year}
  {1981})}\BibitemShut {NoStop}%
\bibitem [{\citenamefont {Belinicher}\ \emph {et~al.}(1982)\citenamefont
  {Belinicher}, \citenamefont {Ivchenko},\ and\ \citenamefont
  {Sturman}}]{BelinicherSturman1982}%
  \BibitemOpen
  \bibfield  {author} {\bibinfo {author} {\bibfnamefont {V.}~\bibnamefont
  {Belinicher}}, \bibinfo {author} {\bibfnamefont {E.}~\bibnamefont
  {Ivchenko}}, \ and\ \bibinfo {author} {\bibfnamefont {B.}~\bibnamefont
  {Sturman}},\ }\href@noop {} {\bibfield  {journal} {\bibinfo  {journal} {Zh.
  Eksp. Teor. Fiz}\ }\textbf {\bibinfo {volume} {83}},\ \bibinfo {pages} {649}
  (\bibinfo {year} {1982})},\ \bibinfo {note} {engl. transl.: Sov. Phys. JETP
  \textbf{56} (2), 359-366 (1982).}\BibitemShut {Stop}%
\bibitem [{\citenamefont {Sodemann}\ and\ \citenamefont
  {Fu}(2015)}]{SodemannFu2015}%
  \BibitemOpen
  \bibfield  {author} {\bibinfo {author} {\bibfnamefont {I.}~\bibnamefont
  {Sodemann}}\ and\ \bibinfo {author} {\bibfnamefont {L.}~\bibnamefont {Fu}},\
  }\href@noop {} {\bibfield  {journal} {\bibinfo  {journal} {Phys. Rev. Lett.}\
  }\textbf {\bibinfo {volume} {115}},\ \bibinfo {pages} {216806} (\bibinfo
  {year} {2015})}\BibitemShut {NoStop}%
\bibitem [{\citenamefont {Zyuzin}\ and\ \citenamefont
  {Zyuzin}(2017)}]{ZyuzinZyuzin2017}%
  \BibitemOpen
  \bibfield  {author} {\bibinfo {author} {\bibfnamefont {A.~A.}\ \bibnamefont
  {Zyuzin}}\ and\ \bibinfo {author} {\bibfnamefont {A.~Y.}\ \bibnamefont
  {Zyuzin}},\ }\href@noop {} {\bibfield  {journal} {\bibinfo  {journal} {Phys.
  Rev. B}\ }\textbf {\bibinfo {volume} {95}},\ \bibinfo {pages} {085127}
  (\bibinfo {year} {2017})}\BibitemShut {NoStop}%
\bibitem [{\citenamefont {Ishizuka}\ \emph {et~al.}(2017)\citenamefont
  {Ishizuka}, \citenamefont {Hayata}, \citenamefont {Ueda},\ and\ \citenamefont
  {Nagaosa}}]{IshizukaNagaosa2017}%
  \BibitemOpen
  \bibfield  {author} {\bibinfo {author} {\bibfnamefont {H.}~\bibnamefont
  {Ishizuka}}, \bibinfo {author} {\bibfnamefont {T.}~\bibnamefont {Hayata}},
  \bibinfo {author} {\bibfnamefont {M.}~\bibnamefont {Ueda}}, \ and\ \bibinfo
  {author} {\bibfnamefont {N.}~\bibnamefont {Nagaosa}},\ }\href@noop {}
  {\bibfield  {journal} {\bibinfo  {journal} {arxiv preprint arXiv:1702.01450}\
  } (\bibinfo {year} {2017})}\BibitemShut {NoStop}%
\bibitem [{\citenamefont {Shapourian}\ and\ \citenamefont
  {Hughes}(2016)}]{ShapourianHughes2016}%
  \BibitemOpen
  \bibfield  {author} {\bibinfo {author} {\bibfnamefont {H.}~\bibnamefont
  {Shapourian}}\ and\ \bibinfo {author} {\bibfnamefont {T.~L.}\ \bibnamefont
  {Hughes}},\ }\href@noop {} {\bibfield  {journal} {\bibinfo  {journal} {Phys.
  Rev. B}\ }\textbf {\bibinfo {volume} {93}},\ \bibinfo {pages} {075108}
  (\bibinfo {year} {2016})}\BibitemShut {NoStop}%
\bibitem{DetassisGrubinskas2017}
F.~Detassis, L.~Fritz, S.~Grubinskas, arXiv preprint arXiv:1703.02425 (2017).  
\bibitem{LyandaGellerAndreev2015}
Y.~B.~Lyanda-Geller, S.~Li, and A.~V.~Andreev, Phys. Rev. B \textbf{92}, 241406(R) (2015).
\bibitem{DeyoSpivak2009}
E.~Deyo, L.~E.~Golub, E.~L.~Ivchenko, B.~Spivak, arXiv preprint arXiv:0904.1917 (2009).
\bibitem{SinitsynSinova2007}
N.~A.~Sinitsyn, A.~H.~MacDonald, T.~Jungwirth, V.~K.~Dugaev, and J.~Sinova, Phys. Rev. B \textbf{75}, 045315 (2007).
\bibitem{AdoTitov2015}
I.~A.~Ado, I.~A.~Dmitriev, P.~M.~Ostrovsky and M.~Titov, Europhys. Lett., \textbf{111}, (2015).
\bibitem{KoenigLevchenko2016}
E. J. K\"{o}nig, P. M. Ostrovsky, M. Dzero, and A. Levchenko, Phys. Rev. B \textbf{94} (2016).
\bibitem{TrescherBergholtz2015}
M.~Trescher, B.~Sbierski, P.~W.~Brouwer, E.~J.~Bergholtz, Phys.~Rev.~B \textbf{91}, 115135 (2015).
\bibitem{Carbotte2016} 
P. Carbotte, Phys. Rev. B \textbf{94}, 165111 (2016).
\bibitem{SteinerPesin2017}
J.~F.~Steiner, A.~V.~Andreev, D.~A.~Pesin, arXiv preprint arXiv:1704.04258 (2017).
\bibitem{YesilyurtJalil2017}
C.~Yesilyurt, S.~G.~Tan, G.~Liang, and M.~B.~A.~Jalil, arXiv preprint 	arXiv:1701.01259 (2017).
\bibitem{MaPesin2015}
J.~Ma and D.~A.~Pesin, Phys. Rev. B \textbf{92}, 235205 (2015).
\bibitem{RouPesin2017}
J. Rou, C. Sahin, J. Ma, D. A. Pesin, preprint arXiv:1705.02367.
\bibitem [{\citenamefont {Burkov}\ and\ \citenamefont
  {Balents}(2011)}]{BurkovBalents2011}%
  \BibitemOpen
  \bibfield  {author} {\bibinfo {author} {\bibfnamefont {A.~A.}\ \bibnamefont
  {Burkov}}\ and\ \bibinfo {author} {\bibfnamefont {L.}~\bibnamefont
  {Balents}},\ }\href@noop {} {\bibfield  {journal} {\bibinfo  {journal} {Phys.
  Rev. Lett.}\ }\textbf {\bibinfo {volume} {107}},\ \bibinfo {pages} {127205}
  (\bibinfo {year} {2011})}\BibitemShut {NoStop}%
\bibitem{BulmashQi2014}
D.~Bulmash, C.-X.~Liu, and X.-L.~Qi, Phys.~Rev.~B \textbf{89}, 081106(R) (2014).
\bibitem [{\citenamefont {Wang}\ \emph {et~al.}(2016)\citenamefont {Wang},
  \citenamefont {Vergniory}, \citenamefont {Kushwaha}, \citenamefont
  {Hirschberger}, \citenamefont {Chulkov}, \citenamefont {Ernst}, \citenamefont
  {Ong}, \citenamefont {Cava},\ and\ \citenamefont
  {Bernevig}}]{WangBernevig2016}%
  \BibitemOpen
  \bibfield  {author} {\bibinfo {author} {\bibfnamefont {Z.}~\bibnamefont
  {Wang}}, \bibinfo {author} {\bibfnamefont {M.~G.}\ \bibnamefont {Vergniory}},
  \bibinfo {author} {\bibfnamefont {S.}~\bibnamefont {Kushwaha}}, \bibinfo
  {author} {\bibfnamefont {M.}~\bibnamefont {Hirschberger}}, \bibinfo {author}
  {\bibfnamefont {E.~V.}\ \bibnamefont {Chulkov}}, \bibinfo {author}
  {\bibfnamefont {A.}~\bibnamefont {Ernst}}, \bibinfo {author} {\bibfnamefont
  {N.~P.}\ \bibnamefont {Ong}}, \bibinfo {author} {\bibfnamefont {R.~J.}\
  \bibnamefont {Cava}}, \ and\ \bibinfo {author} {\bibfnamefont {B.~A.}\
  \bibnamefont {Bernevig}},\ }\href@noop {} {\bibfield  {journal} {\bibinfo
  {journal} {Phys. Rev. Lett.}\ }\textbf {\bibinfo {volume} {117}},\ \bibinfo
  {pages} {236401} (\bibinfo {year} {2016})}\BibitemShut {NoStop}%
\bibitem [{\citenamefont {Patankar}()}]{Patankar2017}%
  \BibitemOpen
  \bibfield  {author} {\bibinfo {author} {\bibfnamefont {S.}~\bibnamefont
  {Patankar}},\ }\href@noop {} {\enquote {\bibinfo {title} {Asymmetric
  scattering induced photocurrent in $\text{TaAs}$},}\ }\bibinfo {note} {Talk
  at the 2017 APS March meeting announced as ``R44:13: Terahertz nonlinear
  optical response from transition metal monopnictide Weyl semimetal
  TaA''.}\BibitemShut {Stop}%
\bibitem [{\citenamefont {Nagaosa}\ \emph {et~al.}(2010)\citenamefont
  {Nagaosa}, \citenamefont {Onoda}, \citenamefont {MacDonald},\ and\
  \citenamefont {Ong}}]{NagaosaOng2010}%
  \BibitemOpen
  \bibfield  {author} {\bibinfo {author} {\bibfnamefont {N.}~\bibnamefont
  {Nagaosa}}, \bibinfo {author} {\bibfnamefont {S.}~\bibnamefont {Onoda}},
  \bibinfo {author} {\bibfnamefont {A.~H.}\ \bibnamefont {MacDonald}}, \ and\
  \bibinfo {author} {\bibfnamefont {N.~P.}\ \bibnamefont {Ong}},\ }\href@noop
  {} {\bibfield  {journal} {\bibinfo  {journal} {Reviews of Modern Physics}\
  }\textbf {\bibinfo {volume} {82}},\ \bibinfo {pages} {1539} (\bibinfo {year}
  {2010})}\BibitemShut {NoStop}%
\end{thebibliography}
\end{document}